\documentclass[twocolumn,twocolappendix]{aastex6}
\pdfoutput=1
\usepackage{amsmath}
\usepackage{natbib}
\usepackage{graphicx}

\shorttitle{Modeling Beta Pic Dust from Optical to Radio}
\shortauthors{Ballering et al.}
\slugcomment{Accepted for publication in ApJ}

\begin{document}

\title{A COMPREHENSIVE DUST MODEL APPLIED TO THE RESOLVED BETA PICTORIS DEBRIS DISK \\ FROM OPTICAL TO RADIO WAVELENGTHS}

\author{Nicholas P. Ballering, Kate Y. L. Su, George H. Rieke, Andr\'as G\'asp\'ar}
\affil{Steward Observatory, University of Arizona, 933 North Cherry Avenue, Tucson, AZ 85721, USA}
\email{ballerin@email.arizona.edu}

\begin{abstract}
We investigate whether varying the dust composition (described by the optical constants) can solve a persistent problem in debris disk modeling---the inability to fit the thermal emission without over-predicting the scattered light. We model five images of the $\beta$ Pictoris disk: two in scattered light from \textit{HST}/STIS at 0.58 $\micron$ and \textit{HST}/WFC3 at 1.16 $\micron$, and three in thermal emission from \textit{Spitzer}/MIPS at 24 $\micron$, \textit{Herschel}/PACS at 70 $\micron$, and ALMA at 870 $\micron$. The WFC3 and MIPS data are published here for the first time. We focus our modeling on the outer part of this disk, consisting of a parent body ring and a halo of small grains. First, we confirm that a model using astronomical silicates cannot simultaneously fit the thermal and scattered light data. Next, we use a simple, generic function for the optical constants to show that varying the dust composition can improve the fit substantially. Finally, we model the dust as a mixture of the most plausible debris constituents: astronomical silicates, water ice, organic refractory material, and vacuum. We achieve a good fit to all datasets with grains composed predominantly of silicates and organics, while ice and vacuum are, at most, present in small amounts. This composition is similar to one derived from previous work on the HR 4796A disk. Our model also fits the thermal SED, scattered light colors, and high-resolution mid-IR data from T-ReCS for this disk. Additionally, we show that sub-blowout grains are a necessary component of the halo.
\end{abstract}


\section{Introduction}
\label{sec:introduction}

Debris disks are the circumstellar material that remains in planetary systems after the giant planets have formed and protoplanetary disks have dispersed, and they provide a unique opportunity to study planetary systems over a large range of orbital scales. The presence of a debris disk confirms that the planet formation process has progressed at least to the formation of planetesimals. The locations of debris disks reveal the architectures of planetary systems, as planets sculpt and clear the debris material \citep[e.g.][]{wyatt1999,moromartin2005,quillen2006,rodigas2014b}. The frequency and brightness of debris disks versus stellar age informs our understanding of the evolution of planetary systems \citep[e.g.][]{rieke2005,su2006,sierchio2014}. Finally, the composition of debris disks---the focus of this study---provides insight into the composition of planetesimals, a critical parameter in understanding their roles in planet formation. For recent reviews of debris disks, see \citet{wyatt2008}, \citet{matthews2014}.         

The particles in a debris disk range in size from parent body planetesimals down to the dust created by the collisional processing of the parent bodies; it is the dust that is primarily observable. Fully characterizing a debris disk involves determining three properties about this dust: its spatial distribution, its size distribution, and its composition. While hundreds of debris disks have been studied, most have only been characterized by their spectral energy distributions (SEDs) that result from infrared thermal emission from the dust \citep[e.g.][]{ballering2013,chen2014}. SEDs provide the temperature of the dust, but the temperature of a dust grain depends on its location, size, and composition; thus, temperature alone is not sufficient to characterize a debris disk fully. Resolved images at multiple wavelengths are much more powerful for characterizing debris disks. An image provides an independent measure of the spatial distribution of the dust, while the variation of its brightness with wavelength allows the size distribution and composition of the dust to be constrained \citep[e.g.][]{debes2008,rodigas2015}. Visible and near-infrared images trace starlight that is scattered by the circumstellar dust grains, while mid-infrared to mm-wave images trace the grains' thermal emission. A complete debris disk model would match all of the available data, including scattered light images, thermal images, and the thermal SED. 

Many studies of debris disks to date have had difficulty successfully modeling both the thermal emission and scattered starlight in a self-consistent manner. \citet{Krist2010} imaged the debris disk around HD 207129 in scattered light with the \textit{Hubble Space Telescope} (\textit{HST}), then modeled the thermal SED of the disk while using the image to fix its location. Assuming the dust was composed of astronomical silicates, they obtained a good fit to the SED by varying the grain size parameters. However, their best fit model significantly over-predicted the brightness of the disk in scattered light compared to the \textit{HST} image. In a very similar analysis, \citet{golimowski2011} modeled the HD 92945 debris disk (also assuming astronomical silicates for the dust composition) and found that their model over-predicted the observed scattered light brightness by a factor of five. \citet{lebreton2012} modeled the thermal SED of the HD 181327 debris disk by varying the grain sizes and composition and fixing the dust location from an \textit{HST} scattered light image at 1.1 $\micron$; their best fit model over-predicted the scattered light brightness by a factor of 4.5. For the HD 32297 debris disk, \citet{rodigas2014} found that the best fitting model to the SED by \citet{donaldson2013} was inconsistent with the disk's scattered light brightness. \citet{rodigas2015}, in characterizing the debris disk around HR 4796A, showed that models fit only to the scattered light data matched the thermal emission data very poorly, and vice versa---illustrating the importance of modeling both the scattered light and thermal emission data simultaneously.

The nearby A6V star $\beta$ Pictoris hosts a large, bright, edge-on debris disk that is amenable to imaging at many wavelengths. The disk was discovered with the {\it Infrared Astronomical Satellite} ({\it IRAS}) through its thermal emission (published by \citet{aumann1985}) and subsequently imaged in scattered light by \citet{smith1984}. Since then, the $\beta$ Pic disk has been observed with numerous instruments and analyzed many times to investigate its various properties. However, no model of the disk has yet been assembled to match the latest high-quality images in scattered light and thermal emission. In this study we perform such an analysis, modeling images from the \textit{Hubble}, \textit{Spitzer}, and \textit{Herschel} space telescopes, and the Atacama Large Millimeter/submillimeter Array (ALMA).

The $\beta$ Pic debris disk consists of multiple components at various stellocentric distances. We focused on the outer two components that were spatially resolved in all the images we considered. These components include a belt of parent body planetesimals and a halo of small dust grains generated by the collisional processing of the parent bodies and pushed into eccentric or unbound orbits by the force from stellar radiation \citep{augereau2001}. The parent body belt---traced by sub-mm images that are sensitive to large grains---extends from $\sim$40 au to $\sim$150 au \citep{dent2014}, while the halo---which dominates the scattered light signal---extends to at least $\sim$1800 au \citep{larwood2001}. Nearer the star is a warm debris component detected in the mid-IR \citep[e.g.][]{knacke1993,telesco2005,chen2007,li2012} and scattered light \citep{milli2014,millarblanchaer2015}, and also a very hot dust component detected from near-IR interferometery \citep{defrere2012}. These inner components were unresolved in many of our data sets, so we did not include them in our analysis. In addition to dust, the $\beta$ Pic debris disk also contains a gas component. The spatial distribution of much of the gas coincides with the dust, and this gas is likely produced by collisional vaporization or photodesorption of the dust grains \citep{brandeker2004,roberge2006,cataldi2014,dent2014}.

The morphology of this debris disk is complicated by several asymmetries and substructures \citep{kalas1995,golimowski2006,apai2015}, which likely originate from perturbations by the giant planet located 8 au from the star \citep{mouillet1997,augereau2001,nesvold2015} that was detected by \citet{lagrange2010}. The goal of our study was to better understand the grain properties---rather than the morphology---of this disk, so we did not attempt to reproduce the observed detailed structure. We did, however, account for the overall brightness asymmetry between the NE and SW sides of the disk by modeling them separately.

An outline of this paper is as follows. In \S\ref{sec:star} we summarize the properties of the central star. In \S\ref{sec:data} we present the data we used in this study, including previously unpublished images from \textit{HST} and \textit{Spitzer}. In \S\ref{sec:modeling} we detail our procedure for generating model images. In \S\ref{sec:spatial} we show our derived spatial parameters for the two outer disk components, which we then adopt when modeling the dust composition as we describe in \S\ref{sec:DustComp}. When modeling the composition we first try grains composed of astronomical silicates (\S \ref{sec:100astrosil}), then we use a generic function for the material optical constants (\S \ref{sec:genericconstants}), and finally we use a mixture of astronomical silicates, water ice, refractory organics, and vacuum (\S \ref{sec:knownconstants}). In \S \ref{sec:sed} we check our best fit model against the thermal SED, while comparisons with additional datasets can be found in the Appendix. In \S\ref{sec:discussion} we discuss the broader implications of our results, then we offer a summary and conclusions in \S\ref{sec:conclusion}.

\section{Stellar Properties}
\label{sec:star}

$\beta$ Pic is a 21--24 Myr-old \citep{binks2014,binks2016} A6V star located at a distance of 19.44 pc \citep{vanleeuwen2007} with $M_\star$ = 1.75 $M_\odot$, $T_\star$ = 8200 K, and $L_\star$ = 8.7 $L_\odot$ \citep{crifo1997}. We required a model SED of the star's photosphere both for measuring the excess infrared flux emerging from the debris disk and for determining the incident flux on dust grains when generating models of the scattered light and thermal emission from the disk. We used an ATLAS9 \citep{castelli2004} photosphere model with $T_\star$ = 8000 K, log g = 4.0, and solar metallicity. The spectrum was modeled only out to 160 $\micron$, so we extended it to 10,000 $\micron$ by extrapolating with a Rayleigh-Jeans power-law. The amplitude of the photosphere SED model was set so that integrating under it yielded a total luminosity of 8.7 $L_\odot$, which required $R_\star$ = 1.54 $R_\odot$. Our final SED model agreed well with photometric data of this star in the visible and near-IR.

\section{Data}
\label{sec:data}

We characterized the $\beta$ Pic outer debris disk by modeling five images in different wavelength regimes. Two images were obtained with \textit{HST} and probe scattered light; they were taken with the Space Telescope Imaging Spectrograph (STIS) and the Wide Field Camera 3 (WFC3). The other three images probed thermal emission and were taken with the Multiband Imaging Photometer for \textit{Spitzer} (MIPS; \citealp{rieke2004}) at 24  $\micron$, the \textit{Herschel} Photodetector Array Camera and Spectrometer (PACS; \citealp{poglitsch2010}) at 70 $\micron$, and ALMA at 870 $\micron$.

In the following sections we describe each of our five datasets, providing extra detail for the \textit{HST}/WFC3 and \textit{Spitzer}/MIPS data that are published here for the first time. There were several basic data processing steps that we applied to all of the images. We cropped the images to place the star at the center. We rotated the images to align the mid-plane of the (edge-on) disk horizontally, using the WCS associated with each image and the disk's known position angle (29$^\circ$). We extracted radial profiles by selecting a strip of each image along the mid-plane of the disk and computed the mean value of the pixels at each point along the length of the strip. There is a known asymmetry between the brightness of the NE and SW sides of the disk, so we extracted the profiles of each side separately. The widths of the strips were 25 pixels (1\farcs27), 11 pixels (1\farcs32), 5 pixels (6\farcs225), 5 pixels (8\arcsec), and 5 pixels (0\farcs5) for the STIS, WFC3, MIPS, PACS, and ALMA images, respectively. For details on how we chose these values, see \S\ref{sec:modeling}.

In Table \ref{table:SED} we present a collection of photometry data for the whole disk spanning the range of wavelengths where the thermal radiation is dominated by the outer disk components ($\lambda$ $\gtrsim$ 20 $\micron$). The \textit{Spitzer} Infrared Spectrograph (IRS; \citealp{houck2004}) data on $\beta$ Pic \citep{chen2007} provides a detailed characterization of dust emission features arising mostly from the inner warm component, which is not the focus of this study, so we do not include it in our SED. The contribution to the flux density for the central star is also listed in the table; it is from the model discussed in \S\ref{sec:star}.

\subsection{HST/STIS}

$\beta$ Pic was imaged with the STIS CCD in coronagraphic (50CORON) mode under program GO-12551 (PI: Apai), and the results of these observations were published in \citet{apai2015}. The observing strategy used multiple roll angles, various coronagraphic wedge positions, and dedicated point-spread function (PSF) star observations to achieve very sensitive imaging of the disk in scattered light, following the technique of \citet{schneider2014}. The instrument bandpass is set by the response of the CCD and centered at 0.58 \micron. While these images achieve a small inner working angle, the field of view of the instrument limited the detection of the disk to $r \lesssim 11\arcsec$ (210 au), well inside of its full extent.

\begin{figure}
\plotone{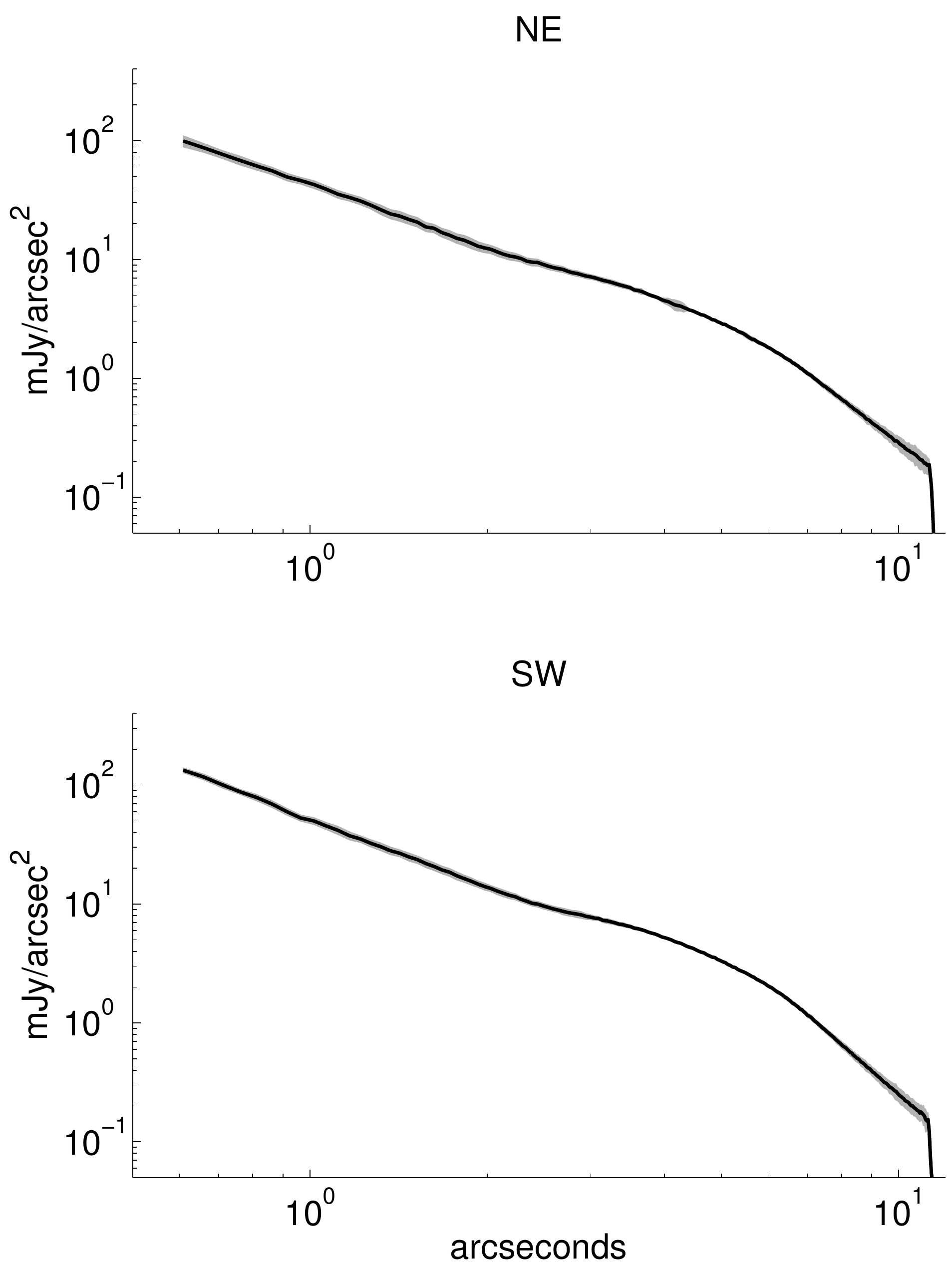}
\caption{\small Radial profiles of the NE and SW sides of the disk at 0.58 $\micron$ from \textit{HST}/STIS. The gray region is the  uncertainty along the profiles. The outer edges of the profiles are truncated by the field of view of STIS.}
\label{fig:STISProfiles}
\end{figure}

We converted the star-subtracted disk image from counts s$^{-1}$ per pixel to mJy arcsec$^{-2}$ using a conversion factor of $4.55\times10^{-7}$ Jy counts$^{-1}$ s and the pixel size of 0\farcs05077 \citep{apai2015}. An image of the uncertainty in each pixel was also provided, and we extracted the radial profile of the uncertainty using the same steps. We combined this in quadrature with a calibration uncertainty of 0.3\% of the signal in the profile. The STIS radial profiles for the NE and SW sides of the disk are show in Figure \ref{fig:STISProfiles}.

\subsection{HST/WFC3}
\label{sec:hst/wfc3}

To detect the full extent of the disk's halo component in scattered light, we needed an image from an instrument with a larger field of view than STIS. We searched the \textit{HST} archive and found previously unpublished observations of $\beta$ Pic with the WFC3 instrument in the IR channel (filter F110W at $\sim$1.16 $\micron$) from program GO-11150 (PI: Graham). We used the pipeline data products that were processed by MultiDrizzle to correct for the geometric distortion inherent in the raw images. No dedicated PSF star observations were taken with this instrument in this program, but images at multiple telescope roll angles were obtained, which we subtracted from each other to remove the light from the central star. Four images were available, each separated by $8^\circ$ of rotation. We converted the images from electron s$^{-1}$ per pixel to mJy arcsec$^{-2}$ using a conversion factor of $6.778\times10^{-8}$ Jy electron$^{-1}$ s (from the FITS file header) and the pixel size of 0\farcs12.

\begin{figure}
\plotone{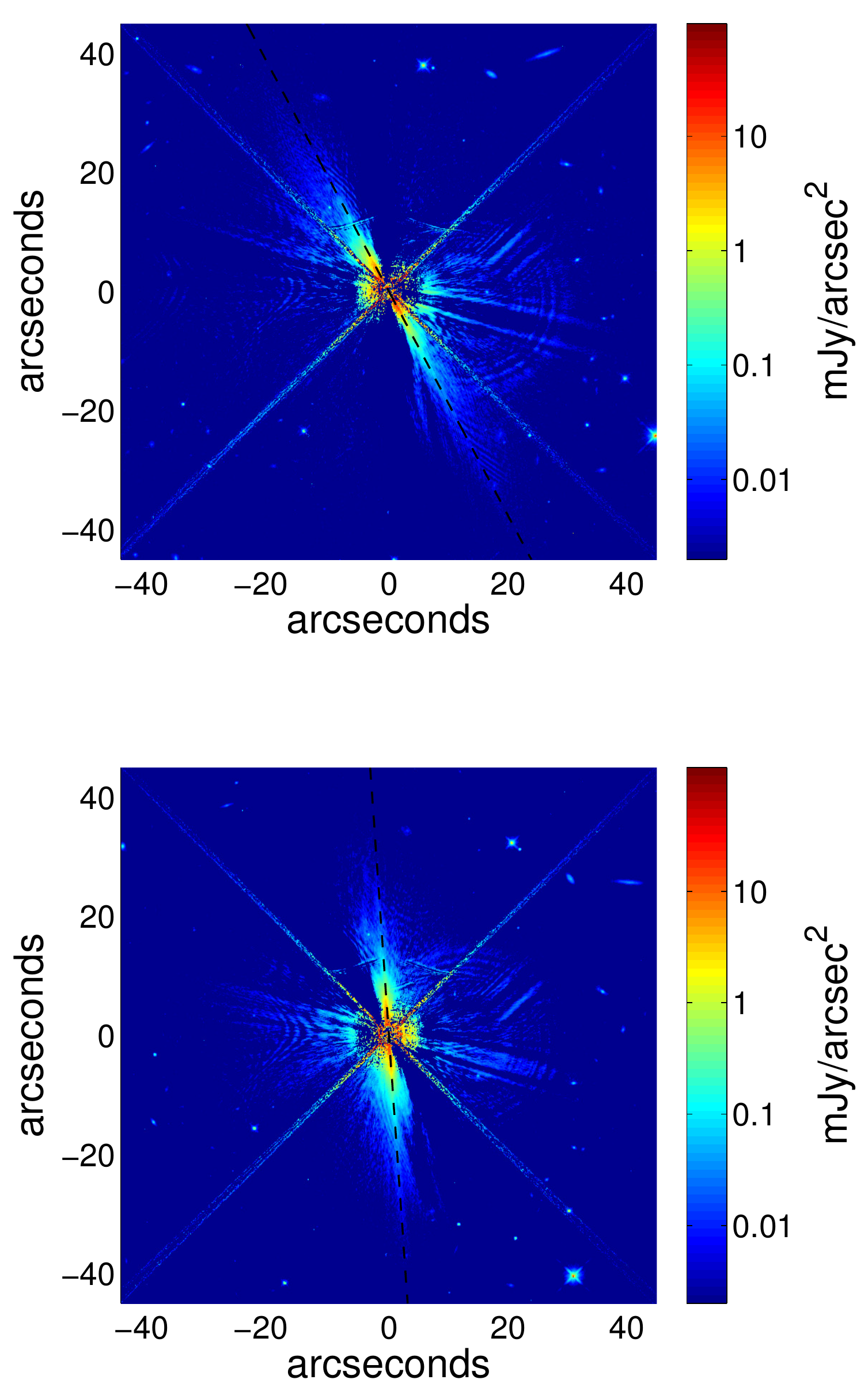}
\caption{\small The difference of two \textit{HST}/WFC3 images with roll angles separated by 24 degrees. The top panel shows the ``positive" image (\texttt{ia1s70031\_drz.fits} $-$ \texttt{ia1s73031\_drz.fits}), where the bottom panel shows the ``negative" image. The black dotted lines locate the midplane of the disk from the two images, along which we generated the radial profiles shown in Figure \ref{fig:WFC3Profiles}. The color is the surface brightness in log scale. In both images the SW side of the disk is up and to the left and the NE side is down and to the right.}
\label{fig:WFC3DiffImage}
\end{figure}

\begin{figure}
\plotone{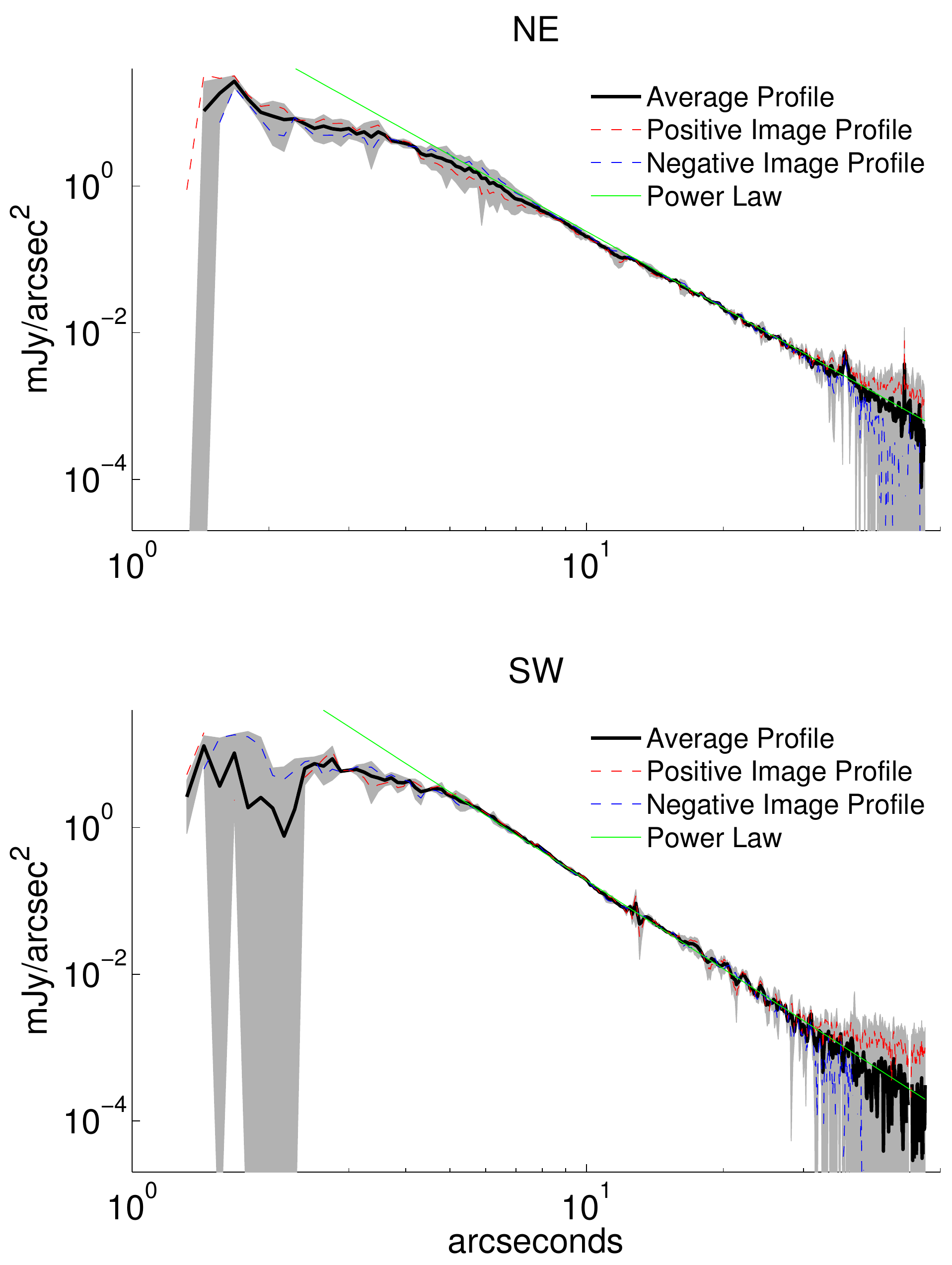}
\caption{\small The NE and SW radial profiles of the disk at 1.16 $\micron$ from the \textit{HST}/WFC3 difference image shown in Figure \ref{fig:WFC3DiffImage}. The final profiles (black lines) are the average of the two profiles from the positive and negative images of the disk (dotted red and blue lines). The gray regions are the uncertainty along the profiles. The green lines are the power law fits to the outer parts ($r>10\arcsec$) of the radial profiles, with indices -3.5 and -4.0 for the NE and SW sides, respectively.}
\label{fig:WFC3Profiles}
\end{figure}

To minimize self-subtraction of the disk signal and extract accurate radial profiles, we opted to use only the two images with the largest difference in rotation angle ($24^\circ$). These were data files \texttt{ia1s70031\_drz.fits} and \texttt{ia1s73031\_drz.fits}. We put each image onto a grid of pixels 10 times smaller than the native pixel size by cubic interpolation with Matlab's \texttt{interp2} function. Prominent PSF diffraction spikes in the images allowed us to accurately align the centers of the images. We subtracted the images and interpolated the difference image back onto the native pixel scale. The result had both a positive and negative disk signal offset by $24^\circ$, and is shown in Figure \ref{fig:WFC3DiffImage}. We extracted radial profiles of both disk images by rotating the difference image to orient each disk horizontally. The final radial profile was the average of these two profiles, and the uncertainty on the final profile was the difference between them. To estimate the amount of flux missing along our radial profiles due to disk self-subtraction, we subtracted two model images (after convolving with the WFC3 PSF, see \S\ref{sec:modeling}) from each other, rotated by $24^\circ$. We then used the result to correct our observed radial profiles. Beyond $r>3\arcsec$, where we perform our fitting (see \S\ref{sec:fittingprocedure}), this correction was relatively small---smaller than our estimated uncertainties. The profiles are shown in Figure \ref{fig:WFC3Profiles}.

\citet{golimowski2006} presented scattered light profiles of this disk measured with the \textit{HST} Advanced Camera for Surveys (ACS). The shape and brightness of our profiles were similar to their results. For a quantitative comparison with the shape of the ACS data, we fit a power law, $S(r) \propto r^\alpha$, to the outer part ($r>10\arcsec$) of our radial profiles. We found $\alpha$ = -3.5 and -4.0 for the NE and SW sides, respectively, which agreed well with the power law fits to the outer part of the (not deconvolved) ACS data, as given in Table 3 of \citet{golimowski2006}. Our power law fits are shown in Figure \ref{fig:WFC3Profiles}.

\subsection{Spitzer/MIPS}

The MIPS observations of $\beta$ Pic were taken under the Spitzer Guaranteed Time Observing Program 90 (PI: M. Werner). The data at all three bands (24, 70, and 160 $\micron$) are published here for the first time. PACS provided a higher spatial resolution image in the far-IR than MIPS (see \S \ref{sec:pacsdata}), so we used the MIPS 70 and 160 $\micron$ data for SED photometry points only (\S \ref{sec:mips70and160data}). The MIPS data were processed using the Data Analysis Tool \citep{gordon2005} for basic reduction. Additional reduction steps, outlined below, were performed on individual exposures which were then mosaicked into one combined image with pixels half the size of the physical pixel scale.

\subsubsection{24 $\micron$}
\label{sec:mips24}

Two sets of 24 $\micron$ observations were obtained. The first set was obtained on 2004 March 20 using 4 sub-pixel cluster positions with a 3 s exposure time and 1 cycle in the large-field photometry Astronomical Observation Template (AOT), resulting in a total of 120 s of integration per pixel. The second set of data was obtained on 2004 April 11 using two large cluster positions with 3 s and 3 cycles in the large-field photometry AOT, resulting in a total of 180 s of integration per pixel. 

At 24 $\micron$, the bright star amplified the ``jailbar" effect, resulting in a striping pattern on each exposure. This striping pattern was removed by subtracting median column offsets in individual exposures. Due to the fine dither pattern in the large-field mode at 24 $\micron$, the bright source (near hard saturation) was exposed to a similar part of the array in sequential exposures, resulting in a potential accumulation of latent images. Since the image latent is flushed out after the bias boost (the onset of an exposure), the data using the first difference in an exposure has the least influence from image latency. To test whether the image latency affected the surface brightness distribution of the central data, we generated two mosaics: one with only the first two differences (short exposure) and the other with the entire data (long exposure), and compared. The difference between the long and short mosaics was within the errors of the observations. The final 24 $\micron$ combined image used only the data obtained at the first epoch due to a more uniform coverage in the mosaic.

PSF subtraction was used to remove the stellar contribution to the image using the brightness of the star predicted by our model of the stellar photosphere, as described in \S\ref{sec:star}. To model the MIPS 24 $\micron$ PSF, we used the \texttt{STinyTim} software with the default throughput curve and assumed a Rayleigh-Jeans source. \citet{engelbracht2007} showed that an \texttt{STinyTim}-produced PSF model could be made more accurate by smoothing it with a 4\farcs41 boxcar function. We achieved this by generating an oversampled model with 0\farcs245 pixels and then smoothed it with an 18 pixel boxcar.

We converted the disk-only image from instrument units of MIPS24 to mJy arcsec$^{-2}$ using a conversion factor of $4.54 \times 10^{-2}$ MJy sr$^{-1}$ MIPS24$^{-1}$ \citep{engelbracht2007}. The pixel scale was subsampled to 1\farcs245, half the physical pixel size. An image of the uncertainty was derived from the square root of the image in instrument units with the same conversion factor applied, and a radial profile of the uncertainty was generated from this image in the same manner as from the image of the signal. This uncertainty profile was combined in quadrature with 4\% calibration uncertainty \citep{engelbracht2007}.

\begin{figure*}
\plotone{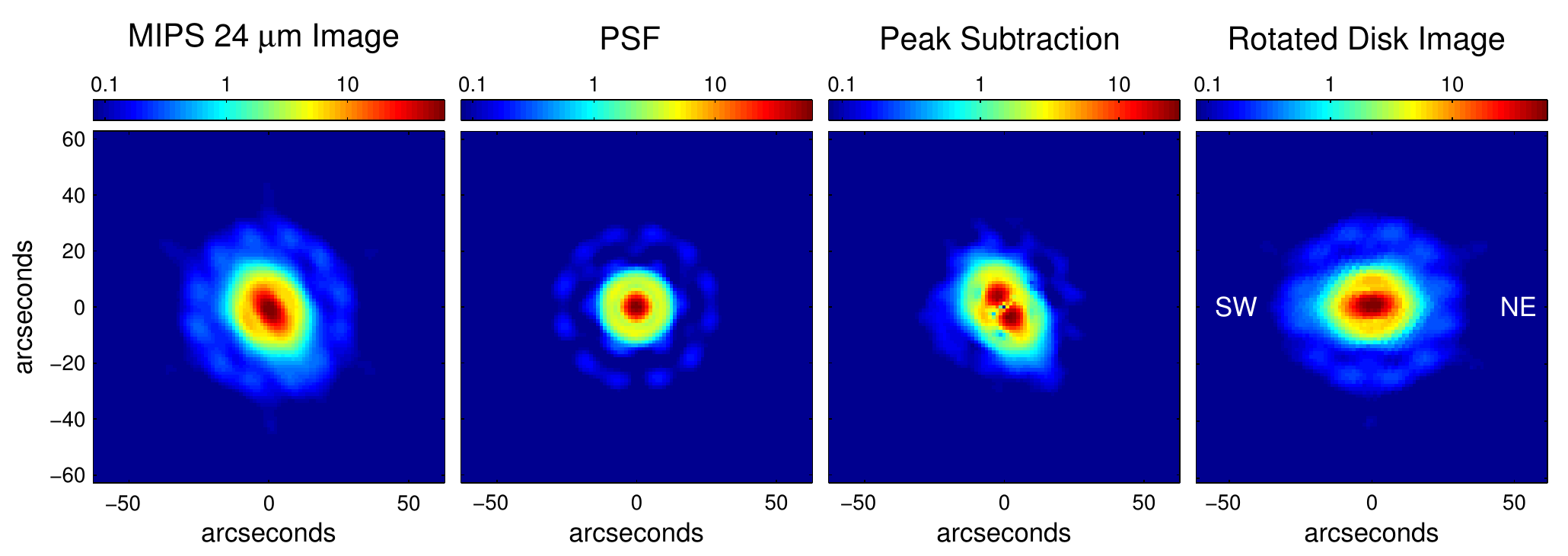}
\caption{\small The first panel shows the \textit{Spitzer}/MIPS image of the $\beta$ Pic system at 24 $\micron$ prior to subtracting the signal from the star. The image is clearly elongated in the NE-SW direction (N is up, E is left), which agrees with the known orientation of the disk. The extension of the observed morphology is clear when comparing the image with the instrument PSF (second panel), which does not exhibit any elongation. We intentionally over-subtracted the PSF model (scaled to match the peak brightness of the observed image) from the observed image, and the result (the third panel) shows residual structure along the orientation of the disk, further confirming that the disk is resolved. In the fourth panel we show the image of the disk with the signal from the star removed and rotated to orient the disk horizontally. The disk accounts for more than 95\% of the total 24 $\micron$ flux from the system, so this disk-only image looks very similar to the image prior to star-subtraction. The color scale in all four images gives the surface brightness (mJy arcsec$^{-2}$) on a log scale.}
\label{fig:MIPS24Image}
\end{figure*}

\begin{figure}
\plotone{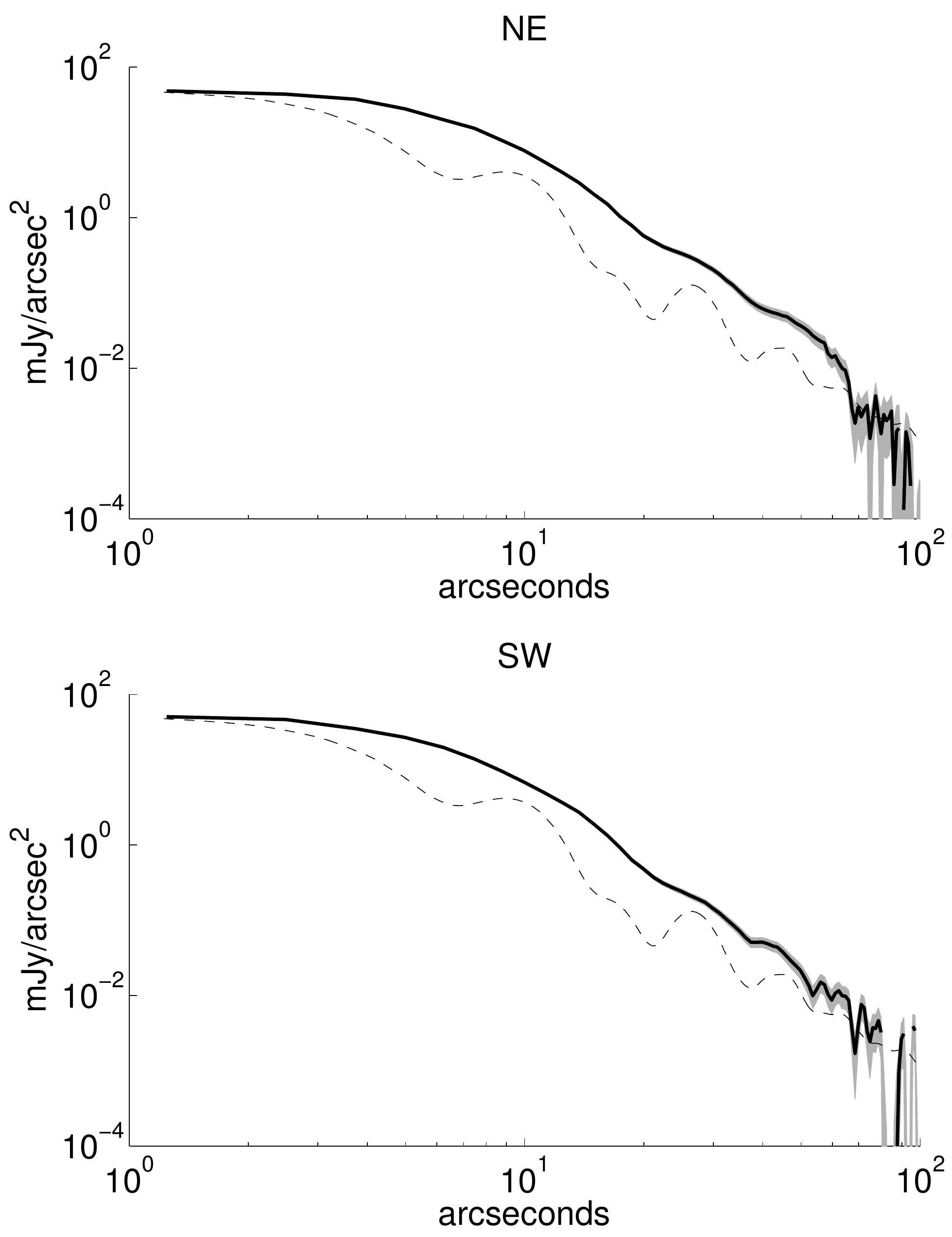}
\caption{\small The radial profiles of the disk at 24 $\micron$. The gray region is the  uncertainty along the profiles. The dashed lines shown the profile of the instrument PSF, scaled to the same peak value as the data's profile.}
\label{fig:MIPS24Profiles}
\end{figure}

Figure \ref{fig:MIPS24Image} shows the MIPS 24 $\micron$ image (first panel), the model PSF (second panel), and the residuals after intentionally over-subtracting the PSF (scaled to the peak brightness of the image) to clearly demonstrate that the disk was resolved by these observations (third panel). In the fourth panel of Figure \ref{fig:MIPS24Image} we show the image of the disk with the signal from the star removed, from which we generated the profiles used for our analysis (these are shown in Figure \ref{fig:MIPS24Profiles}). The first and fourth panels are quite similar because the disk accounts for more than 95\% of the total 24 $\micron$ flux from the system.

In addition to radial profiles, we also measured the total flux density at 24 $\micron$. Before color correction, this was 7.45 Jy using a circular aperture with a radius of 81$\arcsec$ (the maximum flux in the encircled energy method). We applied a color correction of 1.056 (for a blackbody with temperature of 100 K) to only the disk flux (7.13 Jy after subtracting the expected stellar photospheric contribution of 318 mJy), yielding 7.53 Jy for the color-corrected disk flux and 7.85 Jy for the color-corrected total flux. We assumed a 5\% uncertainty on this measurement. Although the total 24 $\micron$ flux exceeded the saturation limit ($\sim$6 Jy for a point source at 3 s exposures), the data were not (although close to) saturated because of the extended structure. At this flux level, there was no significant ($\sim$0.3\%) flux nonlinearity \citep{engelbracht2007}.

\subsubsection{70 and 160 $\micron$}
\label{sec:mips70and160data}

Two sets of 70 $\micron$ observations were obtained. The first set was obtained on 2004 April 12. Unfortunately the disk orientation was along the column direction of the Ge:Ga detector, resulting in much lower sensitivity in the extended disk region. The second set was obtained on 2005 April 4 using 3 cluster positions with 10 s exposure times and 1 cycle in the large-field photometry AOT (a total exposure of $\sim$600 s per pixel). The 160 $\micron$ observation was performed on 2004 February 21 using 7 cluster positions each with 3 s exposure times and 3 cycles in the large-field photometry AOT, resulting in a total of 45 s per pixel.

The 70 $\micron$ data reduction followed the steps recommended by \citet{gordon2007} using time filtering with the source region masked out to avoid filtering out the signal. Several region sizes were tried, and an ellipse with a semimajor radius of 116$\arcsec$ and a semiminor radius of 74$\arcsec$ along the disk midplane (roughly covering the area of the 1-$\sigma$ detection boundary in the final mosaic) gave a minimum value in background variation. The final 70 $\micron$ mosaic used only the data obtained with 10 s exposures (second epoch).

No special steps were performed for the 160 $\micron$ data, and all the exposures were combined based on the WCS information. No leak subtraction was required at 160 $\micron$. It has been shown that the ghost image produced by the 160 $\micron$ filter leakage was less than $\sim$15 times of the photospheric value at 160 $\micron$, whereas the disk was expected to be $\sim$500 times brighter than the expected photospheric value.

The calibration factors we used to transfer the instrument units to physical units were 702 MJy sr$^{-1}$ MIPS70$^{-1}$ \citep{gordon2007} and 41.7 MJy sr$^{-1}$ MIPS160$^{-1}$ \citep{stansberry2007} for the 70 and 160 $\micron$ data, respectively.

At 70 $\micron$, nonlinearity begins to affect the data when a source is brighter than $\sim$1 Jy \citep{gordon2007}, and this nonlinearity becomes apparent for a given pixel when its value is $\geq$0.2 MIPS70 (140.4 MJy sr$^{-1}$). This effect had a significant impact on the observed 70 $\micron$ disk surface brightness distribution as the central 3$\times$3 pixels had values greater than 0.2 MIPS70. We compared the imaging data with the integrated flux in the MIPS-SED data, which was presented by \citet{su2015}. Even though the MIPS-SED observations were obtained with the same detector, the data were unlikely to be in the nonlinear regime because each pixel received less flux due to the dispersive nature of the spectrograph. The MIPS-SED data were taken at three slit positions that covered the NE, center, and SW parts of the disk. Using the non-aperture-corrected MIPS-SED spectra, the integrated flux density in each of the slit positions was 2.3, 7.0, 2.3 Jy for the NE, center, and SW positions, respectively. Using the 70 $\micron$ imaging data, the total flux density within rectangular apertures of 20$\arcsec \times$ 50$\arcsec$ was 2.1, 4.4 and 2.1 Jy for the NE, center, and SW positions, suggesting a $\sim$60\% and $\sim$9\% flux deficit in the central and side regions of the image. After applying a flux nonlinearity correction (K. Gordon et al. 2009, private communication), the corrected 70 $\micron$ image gave 2.5, 7.5, and 2.5 Jy for the NE, center, and SW positions. These values agree with the MIPS-SED data to within 10\%.

We measured the broadband 70 $\micron$ flux of $\beta$ Pic on the flux-nonlinearity-corrected image using the encircled energy method. The total flux density in the 70 $\micron$ band was 16.93 Jy using a circular aperture with radius of 125$\arcsec$ before color correction. The expected stellar photosphere was 32 mJy at 70 $\micron$, suggesting a total disk flux of 18.05 Jy after a color correction of 1.066 (assuming a blackbody of 100 K) with an assumed 10\% error. This value agreed well with the color-corrected IRAS 60 $\micron$ measurement, and was slightly higher than the PACS flux at 70 $\micron$.

The total flux density in the 160 $\micron$ band was 3.6 Jy using an elliptical aperture with semimajor radius of 112$\arcsec$ and semiminor radius of 79$\arcsec$ (covering the area within the 1-$\sigma$ detection level) before color correction. The ghost image due to the 160 $\micron$ filter leak was estimated to contribute $<$3\% of the total flux (less than the calibration error); therefore, no correction was attempted. The total disk flux in the 160 $\micron$ band was 3.65 Jy after a color correction of 1.014 with an assumed 20\% error. This agreed well with ISO point at 170 $\micron$, but was somewhat lower than the PACS flux at 160 $\micron$ (these three measurements agreed to within 3$\sigma$, however).

\subsection{Herschel/PACS at 70 $\micron$}
\label{sec:pacsdata}

\begin{figure}
\plotone{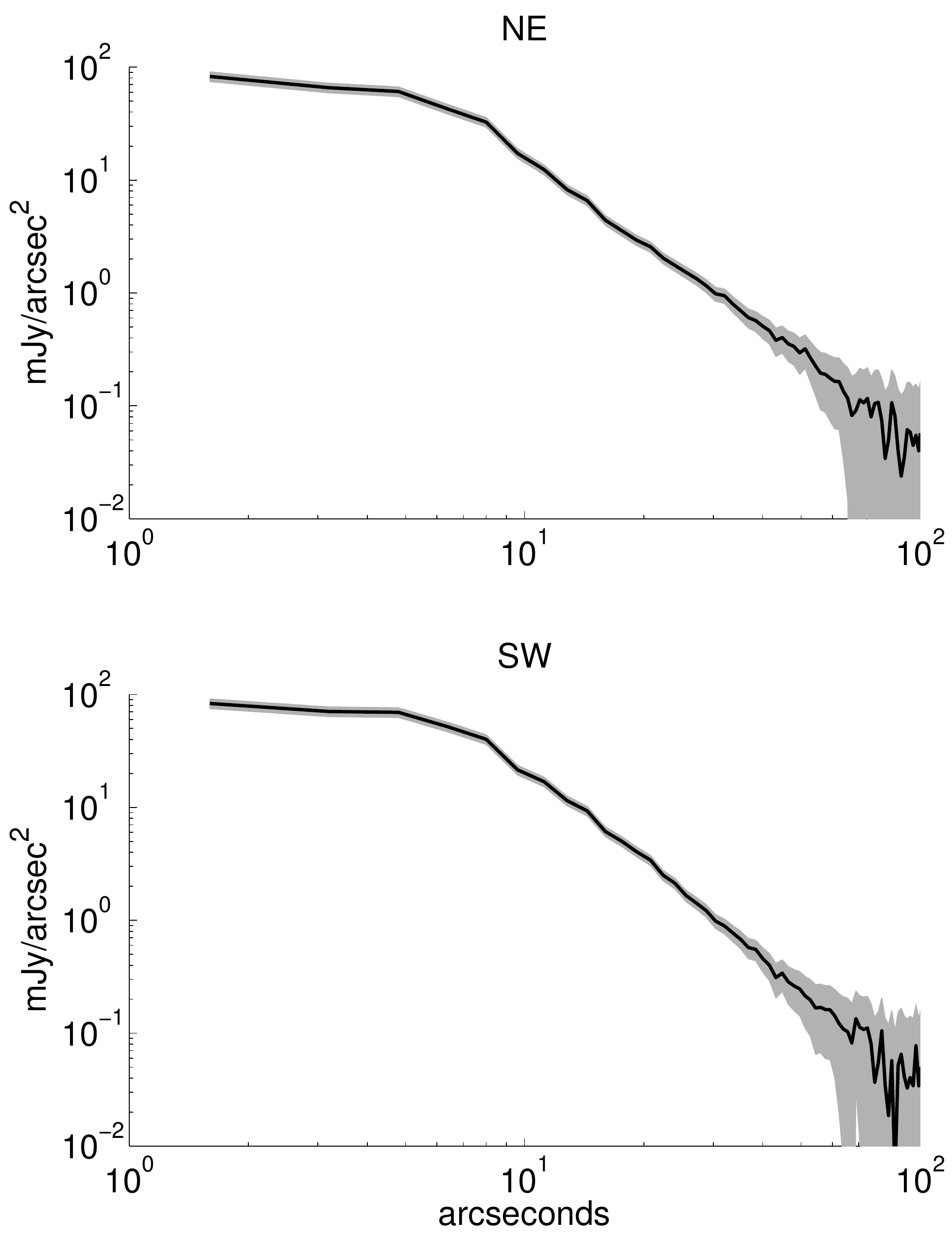}
\caption{\small The radial profiles of the disk from \textit{Herschel}/PACS at 70 $\micron$. The gray region is the uncertainty along the profiles.}
\label{fig:PACSProfiles}
\end{figure}

PACS 70 $\micron$ scan map observations of $\beta$ Pic (PI G. Olofsson, observation IDs 1342186612 and 1342186613) were published by \citet{vandenbussche2010}. We used the Standard Product Generation (SPG) v12.1 level 2.5 corrected MadMap image (a combination of the scan and cross-scan observations) from the Herschel Science Archive. We subtracted a constant background value of 1.3 mJy pixel$^{-1}$ from the image, which was taken to be the median of the pixel values for three regions of the image away from the disk. We converted the units from Jy pixel$^{-1}$ to mJy arcsec$^{-2}$ using the pixel size of 1.6 arcsec. The star contributes a negligible amount of flux compared to the disk, so we did not perform PSF subtraction on this image. The uncertainty on the radial profile was a sum in quadrature of three components: 0.23 mJy pixel$^{-1}$ estimated from the error image supplied by the pipeline processing, 0.19 mJy pixel$^{-1}$ from the median of the standard deviations of the three regions of the original image used to estimate the background, and a 10\% calibration error on the disk profile signal \citep{poglitsch2010}. The radial profiles are shown in Figure \ref{fig:PACSProfiles}. Our profiles agreed with those presented in \citet{vandenbussche2010}.

\subsection{ALMA}

\begin{figure}
\plotone{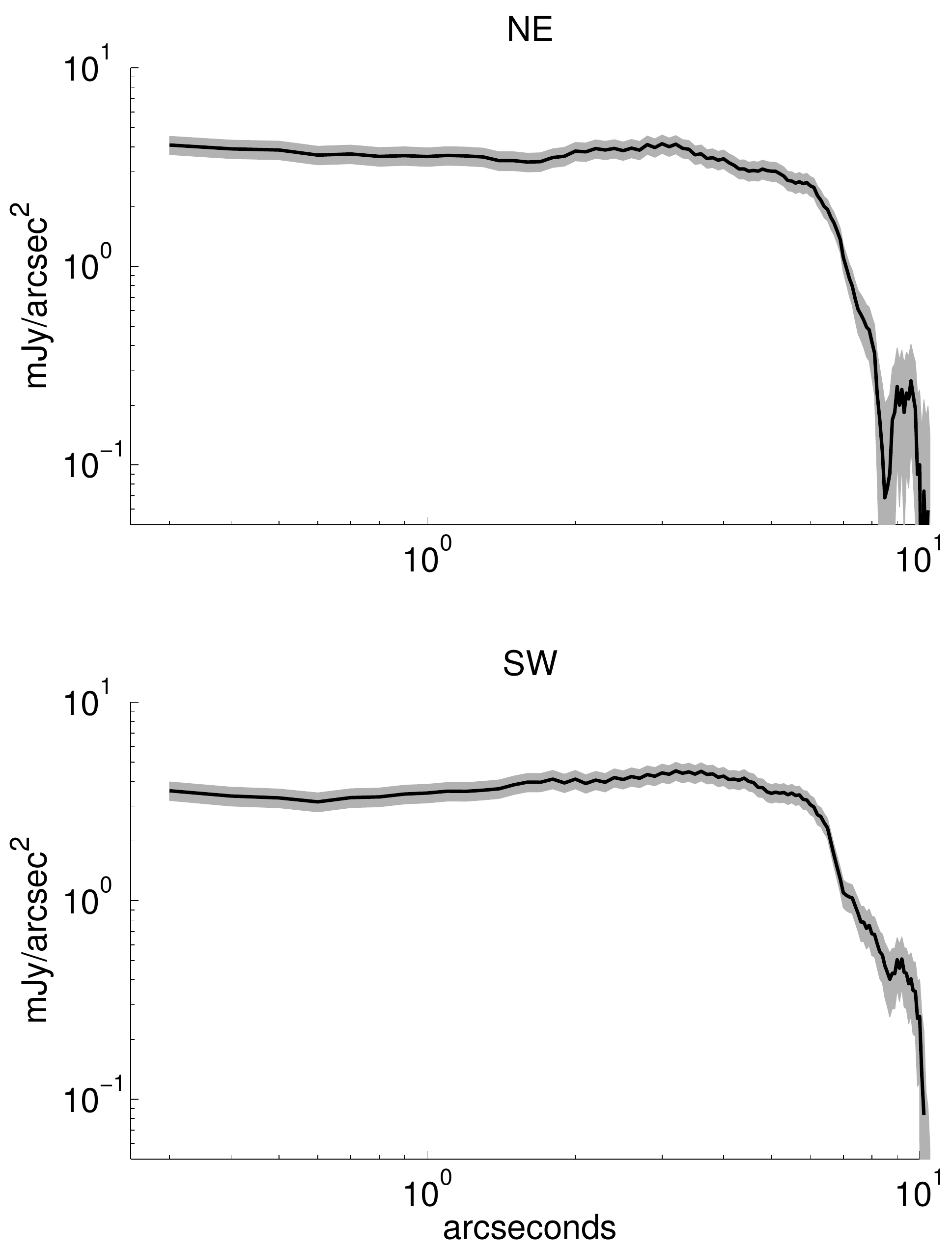}
\caption{\small The radial profiles of the disk from ALMA at 870 $\micron$. The gray region is the uncertainty along the profiles.}
\label{fig:ALMAProfiles}
\end{figure}

We used the ALMA 870 $\micron$ continuum image previously published by \citet{dent2014}. The image had pixels of size 0\farcs1. We converted the image from Jy per beam to mJy arcsec$^{-2}$ using a beam area of $1.133 \times b_\text{maj} \times b_\text{min}$ where $b_\text{maj}=0\farcs709$ and $b_\text{min}=0\farcs556$ are the FWHM of the Gaussian beam major and minor axes, respectively \citep{dent2014}. The star contributes a negligible amount of flux compared to the disk at this wavelength. We created an uncertainty image by combining 0.061 mJy rms uncertainty and 10\% calibration uncertainty in quadrature and then extracted the uncertainty radial profile from this image. The profiles are shown in Figure \ref{fig:ALMAProfiles}.

\begin{deluxetable*}{ccccccc}
\tabletypesize{\scriptsize}
\tablewidth{0pt}
\tablecolumns{7}
\tablecaption{\small Broadband SED Photometry Data at Wavelengths Dominated by the Outer Disk Components \label{table:SED}}
\tablehead{\colhead{$\lambda$ ($\micron$)} & \colhead{Total $F_\nu$ (Jy)} & \colhead{Error $F_\nu$ (Jy)} & \colhead{Star $F_\nu$ (Jy)} & \colhead{Excess $F_\nu$ (Jy)} & \colhead{Instrument} & \colhead{Ref.}}
\startdata
18.30 & 4.316 & 0.432 & 0.498 & 3.818 & TReCS & 2 \\ 
23.67 & 7.847 & 0.392 & 0.318 & 7.529 & MIPS & 1 \\ 
24.60 & 8.807 & 0.881 & 0.296 & 8.511 & TReCS & 2 \\ 
25.00 & 10.200 & 2.000 & 0.288 & 9.912 & ISO & 3 \\ 
25.00 & 10.072 & 1.007 & 0.288 & 9.784 & \textit{IRAS} & 4 \\ 
60.00 & 18.500 & 3.700 & 0.046 & 18.454 & ISO & 3 \\ 
60.00 & 18.930 & 1.893 & 0.046 & 18.884 & \textit{IRAS} & 4 \\ 
70.00 & 16.000 & 0.800 & 0.034 & 15.966 & PACS & 5 \\ 
71.42 & 18.048 & 1.805 & 0.032 & 18.016 & MIPS & 1 \\ 
100.00 & 10.576 & 1.058 & 0.016 & 10.560 & \textit{IRAS} & 4 \\ 
100.00 & 9.800 & 0.500 & 0.016 & 9.784 & PACS & 5 \\ 
155.89 & 3.650 & 0.730 & 0.007 & 3.643 & MIPS & 1 \\ 
160.00 & 5.100 & 0.500 & 0.006 & 5.094 & PACS & 5 \\ 
170.00 & 4.100 & 0.800 & 0.006 & 4.094 & ISO & 3 \\ 
250.00 & 1.900 & 0.100 & 0.003 & 1.897 & SPIRE & 5 \\ 
350.00 & 0.720 & 0.050 & 0.001 & 0.719 & SPIRE & 5 \\ 
500.00 & 0.380 & 0.030 & 0.001 & 0.379 & SPIRE & 5 \\ 
850.00 & 0.058 & 0.006 & 0.000 & 0.058 & SCUBA & 6 \\ 
870.00 & 0.075 & 0.037 & 0.000 & 0.075 & APEX & 7 \\ 
870.00 & 0.060 & 0.006 & 0.000 & 0.060 & ALMA & 8 \\ 
1200.00 & 0.036 & 0.010 & 0.000 & 0.036 & SIMBA & 9 \\ 
\enddata
\tablerefs{
(1) This work
(2) \citet{telesco2005}
(3) \citet{heinrichsen1999}
(4) IRAS Faint Source Catalog (color-corrected values)
(5) \citet{vandenbussche2010}
(6) \citet{holland1998}
(7) \citet{nilsson2009}
(8) \citet{dent2014}
(9) \citet{lisea2003}
}
\end{deluxetable*}

\section{Model Images}
\label{sec:modeling}

In this section we describe how we generated model debris disk images. The specific sets of models are discussed in subsequent sections. As mentioned previously, we focused our modeling effort on the parent body belt and halo components of the disk. We generated model images of these two components separately, and because the disk is optically thin, we could simply sum them together during the fitting process. Each component was modeled as a wedge-shaped\footnote{We set the half-opening angle of the wedge disk to 4$^\circ$, determined by comparing vertical cuts of the WFC3 images with vertical cuts of model images made with a range of opening angles. We assumed the parent body and halo components had that same half-opening angle.} disk extending between inner and outer radial boundaries $r_\text{in}$ and $r_\text{out}$. The number density of grains in the disk varied as a power law with both stellocentric radius and grain size (radius of $a$) as $n(r,a) \propto r^{-p}a^{-q}$, and the grain size distribution was bounded by $a_\text{min}$ and $a_\text{max}$.

We used the code \texttt{dustmap} v.3.1.1 \citep{stark2011} to generate model disk images in both thermal emission and scattered light. The disk geometry was input into \texttt{dustmap} by specifying the Cartesian coordinates of the desired dust distribution. We setup our model space with the star at the origin, the $x$ coordinate to the right, the $y$ coordinate away from the viewer, and the $z$ coordinate up. Our models were seen perfectly edge-on, i.e. $i=90^\circ$ with the midplane of the disk in the $xy$ plane.

We populated each model disk with grains equally spaced in Cartesian space within the defined wedge-shaped disk. The desired pixel size of the model images was $0\farcs05$ in order to be as small as the smallest pixels in our data images (STIS). The spacing of model particles in Cartesian space was set to be the same as the pixel size (0.97 au), with the particles arranged to be located at the center of each pixel in the image plane. We set the field of view of the model image to be a square extending just to the outer edge of the disk. As the number of pixels spanning the field of view must be an integer, the final size of the model pixels differed very slightly from $0\farcs05$.

Each model particle was assigned an ``intensity" value, allowing the particle to represent this number of physical dust grains. We used intensity values to implement the radial variation in grain number density where $r=\sqrt{x^2+y^2}$. \texttt{dustmap} can create a series of model images---each assuming the grains are all of a single size---and then sum the images together with relative ``scaling" values; we used this feature to implement the grain size distribution. For both halo and parent body models, the grain size distribution was sampled with 50 values distributed logarithmically between $a_\text{min}$ and $a_\text{max}$. We scaled each model image to represent a disk with a mass of $10^{-10} M_\odot$ ($3.33 \times 10^{-5} M_\earth$).

The dust composition entered the model via the optical constants of the material, which are the real and imaginary components of its index of refraction, given by $n(\lambda)$ and $k(\lambda)$. From the optical constants, \texttt{dustmap} used Mie theory to compute the absorption and scattering efficiency factors and the scattering phase function of the dust grains when generating model images. We did not calculate the thermal emission contribution to the models at 0.58 and 1.16 $\micron$, nor did we include the scattered light contribution to the model images at 24, 70, and 870 $\micron$; the omitted components contribute negligibly to the total outputs at these wavelengths. \texttt{dustmap} model images were produced in Jy/pixel, which we converted to mJy arcsec$^{-2}$ using the size of the model pixels.

In practice, we generated each model image in pieces to use computer resources more efficiently, taking advantage of symmetries afforded by assuming the disk was perfectly edge-on. We first modeled the region $x>0$, $y>0$, and $z>0$, and also split the $x$ range into two sub-models. Combining these two sub-model images yielded one octant of the disk (the ``back" side of one quadrant of the disk). Thermal emission is radiated isotropically, so for thermal models we doubled this image to model one quadrant of the disk. For the scattered light models, the ``front" side ($y<0$) of that quadrant was modeled separately, then the back and front pieces were added together. Finally, the quadrant model was mirrored over the $x$ and $z$ axes to yield a model image of the full disk.

To compare the model images with the observations, we convolved our models with a model PSF appropriate for each instrument. We used the \texttt{TinyTim} software \citep{Krist2011} to generate model PSFs for the two \textit{HST} images. For the STIS PSF, we used $\lambda$=0.58 $\micron$, a model diameter of 10\arcsec, 0 defocus, and no oversampling. For the WFC3 PSF we used the source spectrum of an A5 star, a model diameter of 10\arcsec, and 0 defocus. We selected the undistorted model, as the data products we used were corrected for the distortion in this instrument. We kept the default orientation of the \textit{HST} model PSFs because our images from these instruments were a product of multiple disk orientations. The MIPS 24 $\micron$ PSF was made using \texttt{STinyTim} as described previously in \S\ref{sec:mips24}, except now assuming the source was a 100 K blackbody. We used PSF models for PACS at 70 $\micron$ derived from observations of Vesta taken with the same scan speed (slow) as our image of $\beta$ Pic.\footnote{Specifically, we used \texttt{psf20\_blu\_10\_vesta\_od160\_ama+63.fits}. Files and documentation for this PSF are found at https://nhscsci.ipac.caltech.edu/sc/index.php/Pacs/PSFs} The ALMA PSF was modeled as an elliptical 2D Gaussian function with FWHM major and minor axes of $0\farcs709$ and $0\farcs556$.

Before convolving with the model images, the MIPS, PACS, and ALMA PSFs were rotated to the same relative orientation with the disk midplane as in the observed data sets. We placed the model image onto the same pixel grid as the PSF model by linear interpolation in log space using MATLAB's \texttt{interp2} function, then we performed the convolution of the two images with MATLAB's \texttt{conv2} function. Finally, the convolved model image was interpolated onto the pixel scale of the observed image, and a radial profile was extracted using the same method as was used for the data (described in \S\ref{sec:data}). Because the models were axisymmetric and we fit to each side of the observed disk separately, we only used half of each model radial profile.

The widths of the strips used to make the radial profiles were chosen to capture the majority of the flux along the midplane of the disk. The disk was unresolved in the vertical direction by MIPS and PACS, so the strip widths were set to $\gtrsim$ the FWHM of these instruments' PSFs. To be symmetric, the strip widths needed to be an odd number of pixels: 5 pixels (6\farcs225) for MIPS and also 5 pixels (8\arcsec) for PACS. The \textit{HST} images did resolve the vertical extent of the disk, and to guide our choice of strip width we examined the model images (which had a vertical extent set to match the observed disk in scattered light). The images were brightest along the midplane and became much fainter above and below the midplane. For the halo images, the full width of the bright region was $\sim$1\farcs3 at $r$ = 10\arcsec. Thus, we used strip widths of 25 pixels (1\farcs27) for STIS and 11 pixels (1\farcs32) for WFC3. As we will show, the ALMA data trace the parent body component. Our parent body models had a bright midplane region with a full width of $\sim$0\farcs5 (at $r$ = 4\arcsec). This is also approximately the resolution of these ALMA data, so we used a strip width of 5 pixels (0\farcs5).

\section{The Dust Spatial Distribution}
\label{sec:spatial}

The first step in our characterization of the $\beta$ Pic debris disk was to model its spatial properties, specifically $r_\text{in}$, $r_\text{out}$, and $p$ for both the halo and parent body components (recall that $n(r,a) \propto r^{-p}a^{-q}$). Because we possessed well-resolved images of the disk, we could determine these parameters independently of the grain properties. Once the spatial parameters were measured, we kept them fixed while modeling the grain properties, as described in the next section.

\begin{figure}
\plotone{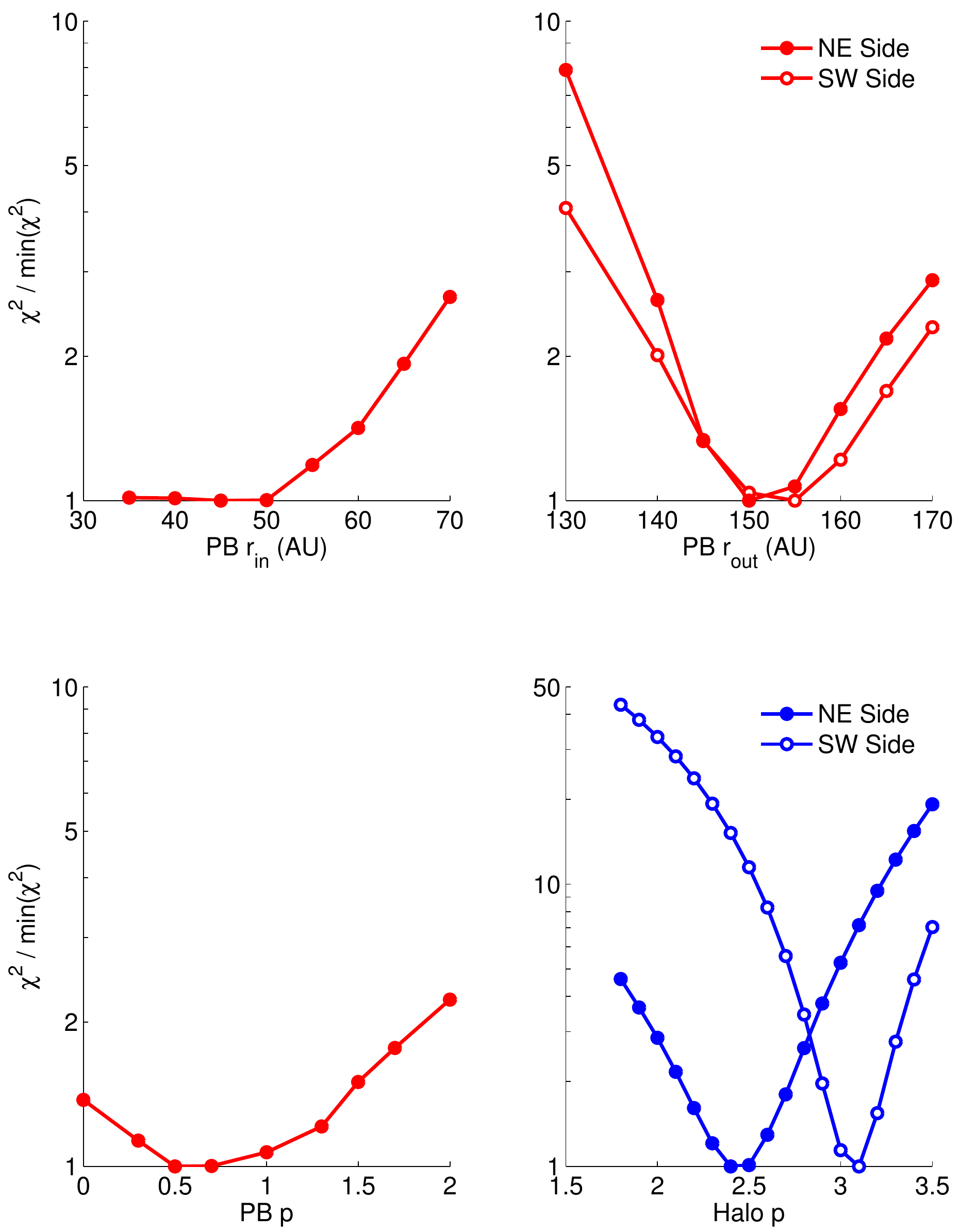}
\caption{\small Constraints on the spatial parameters of the two disk components. The red curves show constraints on the parent body component from the ALMA data. All three spatial parameters were constrained for the NE side, while for the SW side we assumed the same $r_\text{in}$ and $p$ as the NE side but independently constrained its $r_\text{out}$. The blue curves show the constraints on the halo $p$ parameter for the NE and SE sides from the WFC3 data.}
\label{fig:SpatialConstraints}
\end{figure}

At sub-mm wavelengths, the small dust grains that likely constitute the halo component emit very inefficiently and the large grains in the parent body component dominate the signal. Thus, we could constrain the spatial properties of the parent body component by modeling the ALMA image. We performed this fit with a grid search across the parameters of interest, and the results are shown in red in Figure \ref{fig:SpatialConstraints}. We generated models with $r_\text{in}$ ranging from 35 to 70 au, $r_\text{out}$ from 130 to 170 au, and $p$ from 0 to 2.0. For the spatial fitting, the grain properties were fixed at $a_\text{min}$ = 5 $\micron$, $a_\text{max}$ = 5000 $\micron$, $q$ = 3.65, and a composition of astronomical silicates \citep{draine2003}. We also varied the amplitude of the model over a large range of values. For each set of model parameters, we calculated the $\chi^2$ goodness of fit between the model and observed radial profile. For the NE side, the best fit model had $r_\text{in}$ = 45 au, $r_\text{out}$ = 150 au, and $p$ = 0.5. The variable $r_\text{out}$ was fairly well constrained by these data, whereas $r_\text{in}$ and $p$ were not as well constrained. For the SW side, we thus tried using the same $r_\text{in}$ and $p$ as found for the NE side, but allowing $r_\text{out}$ to vary. This yielded a very good fit with $r_\text{out}$ = 155 au. We conclude that the parent body component does not show any prominent asymmetries in terms of these parameters. Our values of $r_\text{in}$ and $r_\text{out}$ agree well with the analysis by \citet{dent2014} who modeled these data with concentric dust annuli (see their Figure 3C).

With the spatial properties of the parent body component fixed, we next addressed the halo. Because the grains in the halo are thought to be generated by collisions in the parent body belt, we used the same $r_\text{in}$ for both components. Unlike the ALMA data, the WFC3, MIPS, and PACS radial profiles showed no sharp truncation---the flux from the disk was simply lost in the noise at the outer edge. The PACS data showed signal to the largest radius, so we used these data to constrain $r_\text{out}$. We found $r_\text{out}\geq$1800 au for both the NE and SE sides, which was consistent with the detection of the disk to 1835 au by \citet{larwood2001}. To measure $p$, we used the WFC3 data because the shape of the WFC3 profiles were not significantly influenced by the instrument PSF. To ensure that we modeled only the halo component, we fit to the portion of the radial profile for $r>8\arcsec$ ($>$155 au). We fixed the grain properties to $a_\text{min}$ = 0.1 $\micron$, $a_\text{max}$ = 5 $\micron$, $q$ = 3.65, and a composition of astronomical silicates. We found best fit $p$ values of 2.4 for the NE side and 3.1 for the SW (see the blue curves in Figure \ref{fig:SpatialConstraints}). 

As a check of our best fit halo $p$ values, we used the relation from \citet{strubbe2006} for an edge-on disk that $\alpha=\gamma-\eta-1$, where $\alpha$ is the observed surface brightness power law exponent, $\gamma$ is the disk surface density power law exponent, and $\eta$ describes the opening of the disk as $h=r^\eta$. For our wedge-shaped models, $\eta=1$. Also, $\gamma=-p+\eta$, so $\alpha=-p-1$. Thus, our measured $\alpha$ values of -3.5 and -4.0 from \S\ref{sec:hst/wfc3} predicted $p$ values of 2.5 and 3.0 for the NE and SW sides, which agreed with what we found from model fitting. According to \citet{strubbe2006}, a collision-dominated halo has $\alpha=-3.5$ while a drag-dominated halo has $\alpha=-4.5$. $\beta$ Pic's NE side agreed with the collision-dominated case, whereas the SW side fell between the two cases.

In summary, we found the following spatial parameters. For the NE side of the disk $r_\text{in}$ = 45 au, the parent body $r_\text{out}$ = 150 au, the parent body $p$ = 0.5, the halo $r_\text{out}$ = 1800 au, and the halo $p$ = 2.4. For the SW side the spatial parameters were the same except that the parent body $r_\text{out}$ = 155 au and the halo $p$ = 3.1.

After successfully modeling the dust composition with a mixture of common materials (\S \ref{sec:knownconstants}), we fit the spatial parameters again, this time using the grain composition and size parameters of that best fit model. The results agreed with spatial parameters we found here when assuming the dust consisted purely of astronomical silicates. 

\section{The Dust Composition}
\label{sec:DustComp}

With the spatial parameters of the halo and parent body components determined, we next constrained the grain sizes and compositions by fitting models to our five images of the outer disk simultaneously. We performed our fitting only on the NE side, then checked if the same dust composition could also fit the SW side data (\S\ref{sec:swcheck}). In Section \ref{sec:100astrosil} we show that the data cannot be reproduced with grains consisting entirely of astronomical silicates. In Section \ref{sec:genericconstants} we find that a relatively good fit to the data can be obtained with a simple parametrized model for the dust optical constants. Then, in Section \ref{sec:knownconstants} we find a good fit to our data with grains consisting of a mixture of common materials and derive significant constraints on the allowed grain composition. We begin, however, in the next section with a description of our model-fitting procedure.

\subsection{Fitting Procedure}
\label{sec:fittingprocedure}

In principle, there were six parameters describing the grain sizes: $a_\text{min}$, $a_\text{max}$, and $q$ for both disk components; however, before fitting we used physical arguments to narrow these to four free parameters. The largest particles in the parent body component were the planetesimals that resupply the dust through collisions. However, the total surface area in these large bodies was small, so their contribution to the observed signal was insignificant. Thus, we set $a_\text{max}$ of the parent body component to 5000 $\micron$, an arbitrary but sufficiently large value so the emission from grains larger than this does not contribute significantly to the total. The dust in the halo consisted of the smallest grains generated by the parent body collisional cascade---the grains small enough to have their orbits perturbed by the stellar radiation force. To model this, we defined the ``transition" grain size, $a_\text{tran}$, as a free parameter and set $a_\text{min}$ of the parent body component and $a_\text{max}$ of the halo component to this value. Therefore, the halo and parent body components overlapped spatially (they had the same $r_\text{in}$) but were segregated by grain size.

In addition to the grain size parameters (halo $a_\text{min}$, $a_\text{tran}$, halo $q$, and PB $q$), the dust masses of each component, $M_\text{PB}$ and $M_\text{halo}$, were also free parameters. Because debris disks are optically thin at all wavelengths, the final radial profile model that we compared to the data was the linear combination of the parent body and halo model profiles. The amplitudes of the two model components were directly proportional to $M_\text{PB}$ and $M_\text{halo}$. Although we fit to only one side of the disk, these masses refer to the total mass of the model disk components (both sides). Finally, there were the free parameters describing the dust composition, which were specific to the analyses described in the following sections.

We performed our fitting with a grid search, populating a chi-squared matrix for each of the five images, with one dimension of the matrix for each free parameter. We then combined these matrices according to
\begin{equation}
\label{eq:combinechisq}
\chi^2 = \frac{1}{5} \displaystyle\sum_{I} \frac{\chi_I^2}{\text{min}\left(\chi_I^2\right)},
\end{equation}
with $I$ representing each of the five images. We normalized the matrix from each image by the $\chi^2$ value of the best fitting model to that image in an attempt to weight the contribution from each of the five images equally. To find the constraints on a given free parameter, we stepped that parameter over its range of values, and at each point we searched the $\chi^2$ matrix for the minimum over every combination of the other parameters.

For the fitting described in the following sections, we fit the radial profile outwards of 3$\arcsec$ because in the WFC3 image the flux measured near the star was more likely to have been artificially reduced due to self-subtraction of the disk. This also minimized the influence of the $r_\text{in}$ spatial parameter on the fits, which was not well constrained. The outer edge of the fitting was specific to each band. 

\subsection{Results with 100\% Astronomical Silicates}
\label{sec:100astrosil}

Many previous studies of debris disks---both analyses of images and SEDs---simply assumed the dust was composed entirely of astronomical silicates \citep[e.g.][]{Krist2010,golimowski2011,ertel2011}. ``Astronomical silicates", however, is not well-defined material, rather it is a set of optical constants resembling silicates that has been optimized to reproduce the ISM UV extinction curve. The latest version of these optical constants is given by \citet{draine2003}. It was, nevertheless, useful to model the $\beta$ Pic disk with these optical constants, as doing so allowed us to compare our results more directly with those from other studies. Furthermore, there may be no clear superior alternative to astronomical silicates. Debris disks almost certainly do have a significant silicate component to their composition, as shown by the detection of the distinctive emission feature at $\sim$10 $\micron$ in the Spitzer/IRS spectra of many debris disks \citep{ballering2014,mittal2015}, including the inner warm component of the $\beta$ Pic disk \citep{knacke1993,chen2007}. However, a precise laboratory analog to the silicates in debris disk dust---and a set of associated optical constants spanning the UV to the mm---is not known.

\begin{figure}
\plotone{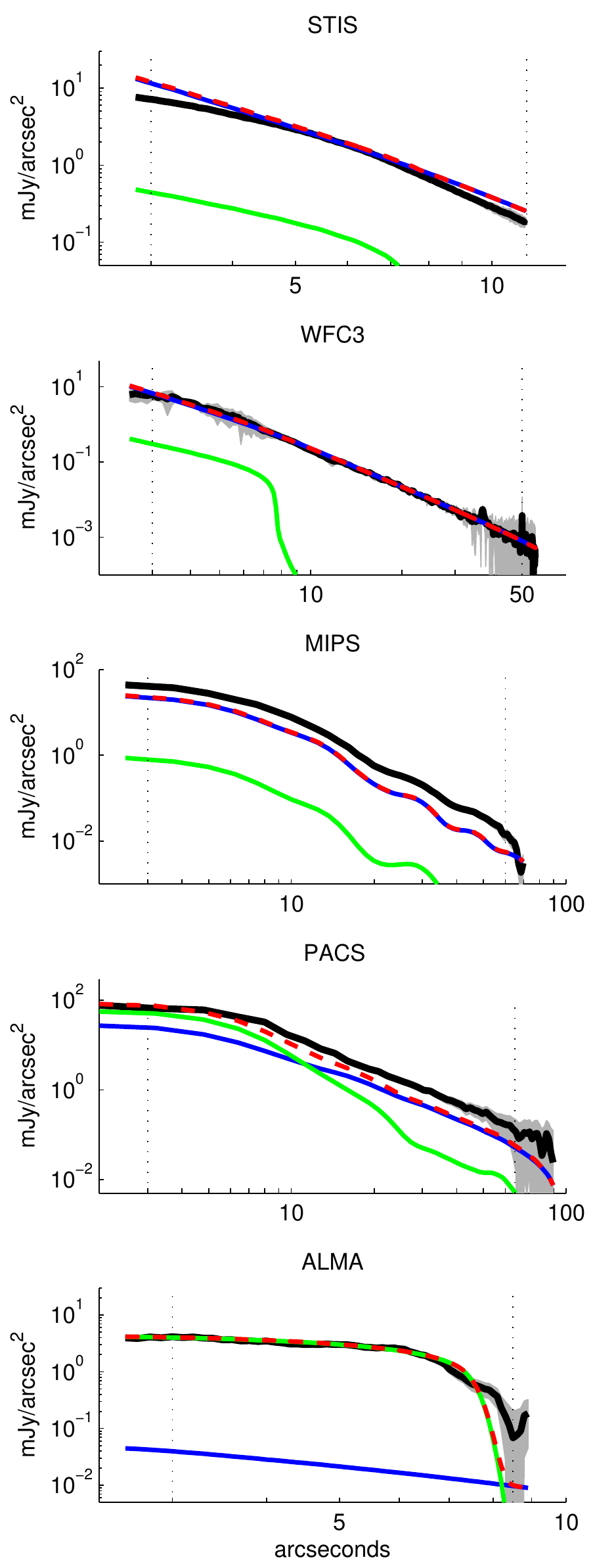}
\caption{\small The best fit model compared with the five data sets for the NE side of the disk, assuming a composition of 100\% astronomical silicates. This illustrates that models with this composition cannot simultaneously fit both the thermal and scattered light data. For example, the model prediction lies above the STIS profile but below those from MIPS and PACS. The black lines are the data, the green lines are the parent body model, the blue lines are the halo model, and the dashed red lines are the total model. The vertical dashed lines show the range of data to which the model was fit. }
\label{fig:allastrosil}
\end{figure}

We performed the fitting while varying the grain size parameters over the following values: halo $q$ = [3, 4], halo $a_\text{min}$ = [0.1, 0.5] $\micron$, and $a_\text{tran}$ = [2, 5] $\micron$. Theoretical examinations of collisional cascades show that $q$ = 3.65 \citep{gaspar2012}, so we adopted this for the parent body component. The best fit was obtained with $a_\text{min}$ = 0.1 $\micron$, $a_\text{tran}$ = 5 $\micron$, and halo $q$ = 3. However, as shown in Figure \ref{fig:allastrosil}, this model did not achieve a good fit to all of the images. Specifically, the halo component (which provided the link between the thermal and scattered data) was fit well to the WFC3 data, but the model was too faint for the MIPS and PACS thermal emission data (and also somewhat too bright at the shorter wavelength STIS scattered light data). This was consistent with the mismatch between the scattered light and thermal emission found in attempts to model other debris disks with astronomical silicates---the model, when fit to the thermal data, was too bright compared to the scattered light observations.        

\subsection{Results with Generic Optical Constants}
\label{sec:genericconstants}

When modeling the dust composition, one is fundamentally manipulating the optical constants, and it is possible that there are degeneracies in this procedure---different mixes of grain compositions might produce similar optical constants and thus similar fits to broad-band data. We therefore start the discussion of optimizing the fit to $\beta$ Pic by considering the optical constants themselves. We generated ``generic" optical constants with only a few free parameters roughly modeled after astronomical silicates. The imaginary component, $k(\lambda)$, of astronomical silicates shows two broad maxima with a trough between, and goes to zero outside the maxima. We modeled this behavior with the piecewise step function
\begin{equation}
\label{eq:generick}
k(\lambda) = \begin{cases} 0 & \lambda < 0.05 \micron \\
k_1 & 0.05 \micron < \lambda < 0.2 \micron \\
k_2 & 0.2 \micron < \lambda < 8 \micron \\
k_3 & 8 \micron < \lambda < 1000 \micron \\
0 & \lambda > 1000 \micron \\
\end{cases}.
\end{equation}
We derived $n(\lambda)$ from $k(\lambda)$ using the Kramers-Kronig relation 
\begin{equation}
\label{eq:kk}
n(\omega) = 1 + \frac{2}{\pi}\int_0^\infty \frac{\Omega k(\Omega)}{\Omega^2 - \omega^2}\,\mathrm{d}\Omega,
\end{equation}
where $\omega = 2\pi c/\lambda$. When evaluating Equation (\ref{eq:kk}) numerically, we avoided the singularity by splitting the integral into two pieces, $\Omega < \omega$ and $\Omega > \omega$, then summed the results. Negative values of $n(\lambda)$ sometimes arose from this procedure at the wavelengths where $k(\lambda)$ was discontinuous; we removed these negative values from the optical constants before passing them to the modeling code. 

\begin{figure}
\plotone{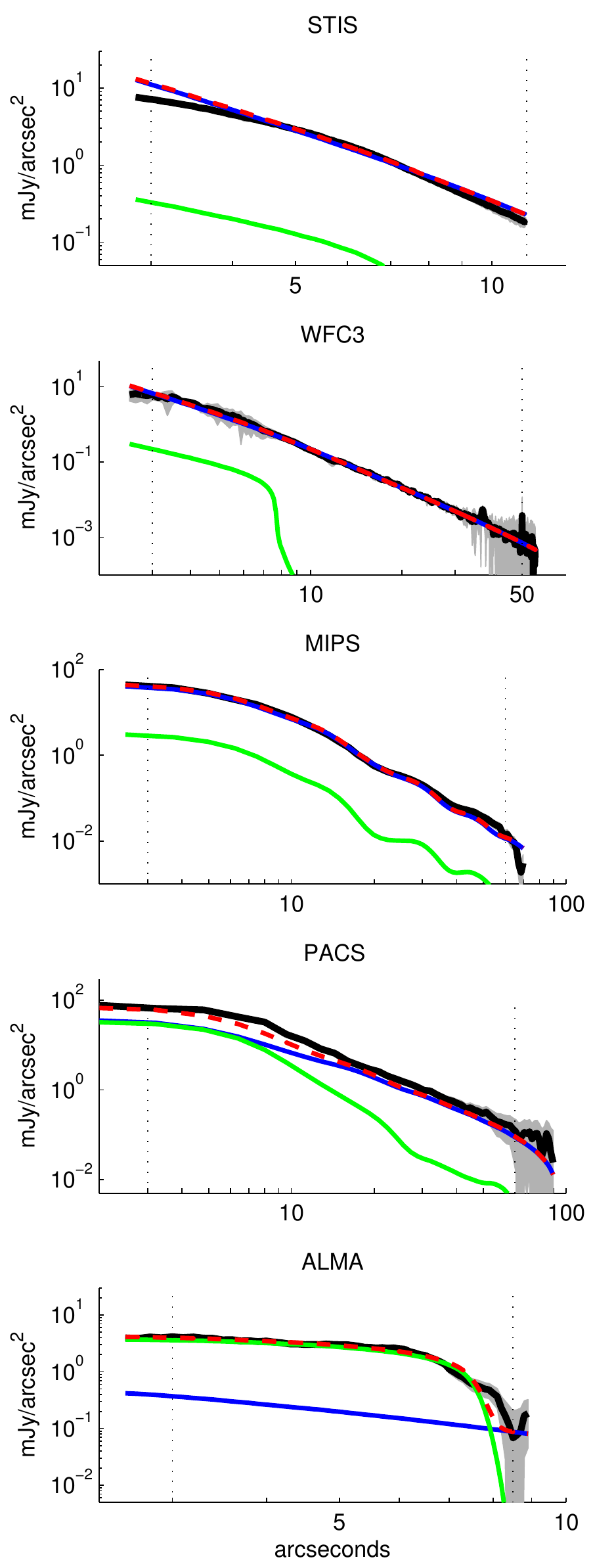}
\caption{\small The best fit model compared to the five data sets using generic optical constants. This shows a much better fit compared to Figure \ref{fig:allastrosil}, illustrating the potential for improving the fitting by modifying the optical constants, even using a very simple model to do so. The black lines are the data, the green lines are the parent body model, the blue lines are the halo model, and the dashed red lines are the total model. The vertical dashed lines show the range of data to which the model was fit.}
\label{fig:genericfitsNE}
\end{figure}

We assumed both components had the same composition, as the grains in the halo are generated from collisions in the parent body belt. We again fit to the five radial profiles, using the grain size parameters of the best model from the previous section (halo $a_\text{min}$ = 0.1 $\micron$, $a_\text{tran}$ = 5 $\micron$, halo $q$ = 3, and PB $q$ = 3.65). We allowed $k_1$, $k_2$, and $k_3$ each to take the values [0.1, 0.6, 1.2]. We found that the best fit model had $k_1$ = 0.6, $k_2$ = 0.1, and $k_3$ = 0.1. The best model is compared with the five data sets in Figure \ref{fig:genericfitsNE}, showing that varying the optical constants can significantly improve the fits, even with a very simple prescription for their form.

\subsection{Results with Mixtures of Common Materials}
\label{sec:knownconstants}

\begin{figure*}
\plotone{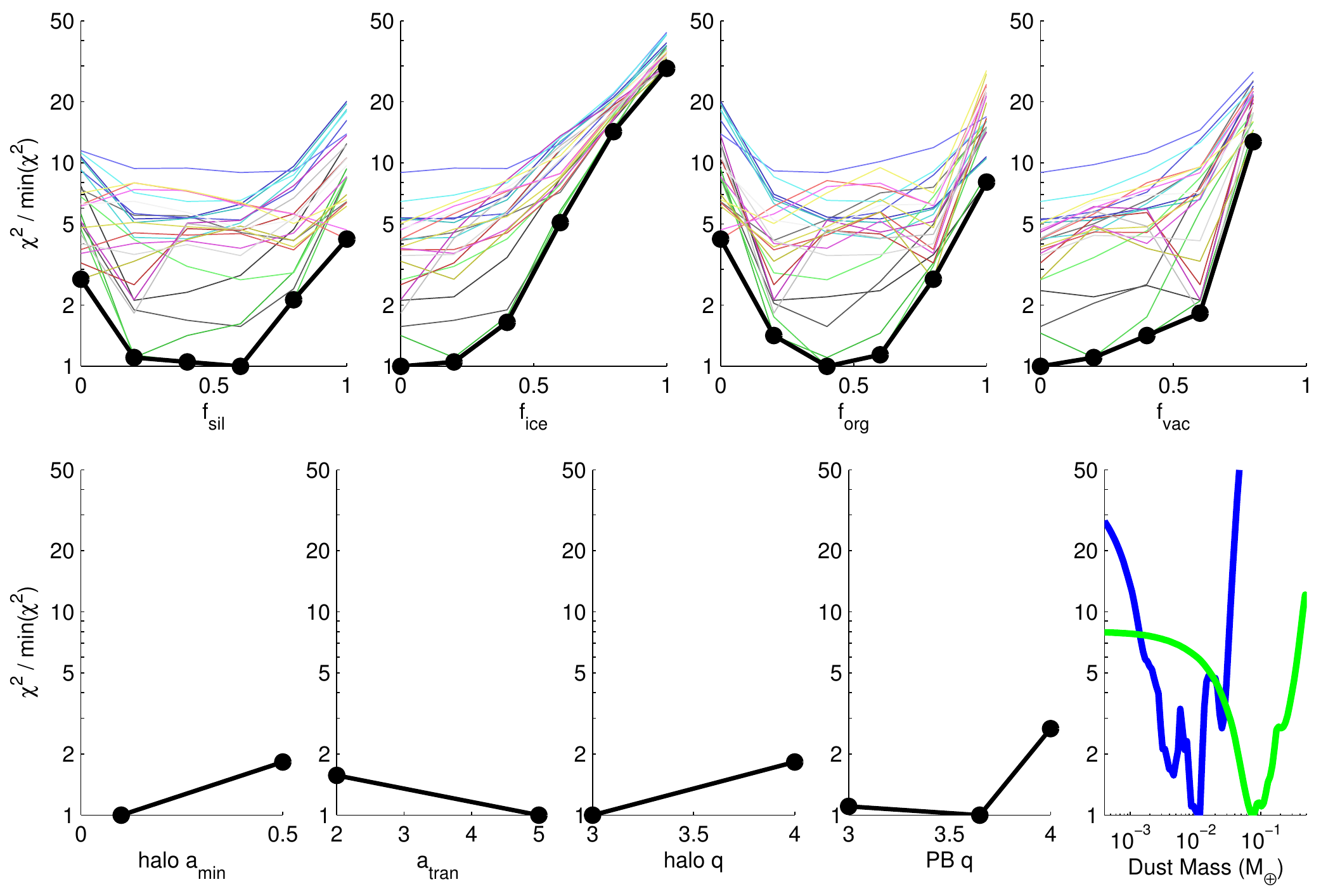}
\caption{\small The $\chi^2$ curves (normalized to the value of the best model) for fitting the dust composition with a mixture of common materials. The thick lines give the best $\chi^2$ values allowing all other parameters to vary. In the dust mass plot, the blue and green curves represent the halo and parent body models, respectively. The 24 thin lines in the top four plots are the curves with each combination of values of the grain size parameters ($a_\text{min}$, $a_\text{tran}$, halo $q$, PB $q$) fixed. This shows that the conclusions about grain composition do not depend strongly on the grain size parameters; that is, a mixture of astronomical silicates and organic refractory material with little to no ice or vacuum is favored regardless of the choice of $a_\text{min}$, $a_\text{tran}$, halo $q$, or PB $q$.}
\label{fig:realcompparams}
\end{figure*}

We now model the disk by mixing the optical constants of known materials. In principle, a broad variety of mixtures of materials might be able to approximate the desired optical constants, so we must use other constraints to guide the assumed grain composition. We kept the number of constituent materials to a minimum while still accounting for the primary types of materials expected. \citet{johnson2012b} simulated the formation of planetesimals in the outer parts of protoplanetary disks of various C/O ratios, redox conditions, and temperatures. While the exact compositions of the resulting planetesimals depended on these disk parameters, the most common materials were always refractory silicates and metals, water ice, and simple carbon-bearing compounds that existed as ices or were trapped in the water ice as clathrates. These carbon-bearing ices and clathrates can be transformed into refractory complex organic material (sometimes called ``ice-tholins") by exposure to UV radiation or cosmic rays \citep{khare1993,mcdonald1996,materese2014}. This processing makes the material darker (lower albedo) and redder. Refractory organics are invoked to explain the low albedo and red color of some objects in the outer solar system \citep{cruikshank2005}.

We therefore proceeded with four materials: astronomical silicates with optical constants from \citet{draine2003}, water ice with optical constants from \citet{li1998}, refractory organic material with optical constants from \citet{li1997}, and vacuum (to model grain porosity) with $n$=1 and $k$=0 at all wavelengths. There are multiple sets of optical constants available for both water ice and organics (e.g. see Table 4 of \citet{rodigas2015}). We selected these specific constants because they had been used previously by \citet{li1998} to model $\beta$ Pic's mid-IR spectral features. The grain densities were 2.7 g/cm$^3$ for the astronomical silicates (as is commonly assumed), and 1.2 and 1.8 g/cm$^3$ for the ice and organics, respectively \citep{li1998}.

The mixing of these component materials was parametrized by the volume fraction of each material, $f_\text{sil}$, $f_\text{ice}$, $f_\text{org}$, and $f_\text{vac}$ with the sum of these fractions equal to unity. We derived the composite optical constants using the Bruggeman mixing rule,
\begin{equation}
\label{eq:bruggeman}
\displaystyle\sum_{j} f_j \frac{\epsilon_j - \epsilon_\text{av}}{\epsilon_j + 2\epsilon_\text{av}} = 0,
\end{equation}
where $\epsilon=\epsilon_1+i\epsilon_2$ is the complex dielectric constant, $\epsilon_\text{av}$ is the dielectric constant of the combined material, and $j$ represents the materials to be combined. The dielectric constant is related to the optical constants by
\begin{equation}
\epsilon_1=n^2-k^2,
\end{equation}
\begin{equation}
\epsilon_2=2nk,
\end{equation}
\begin{equation}
n=\frac{1}{\sqrt{2}}\sqrt{\sqrt{\epsilon_1^2+\epsilon_2^2}+\epsilon_1},
\end{equation}
and
\begin{equation}
k=\frac{1}{\sqrt{2}}\sqrt{\sqrt{\epsilon_1^2+\epsilon_2^2}-\epsilon_1}.
\end{equation}
Note that $\epsilon_1$, $\epsilon_2$, $n$, and $k$ are functions of wavelength.

Fitting the grain properties with these optical constants involved 10 free-parameters: the halo's $a_\text{min}$, $a_\text{tran}$, the halo $q$, the PB $q$, $f_\text{sil}$, $f_\text{ice}$, $f_\text{org}$, $f_\text{vac}$, $M_\text{PB}$, and $M_\text{halo}$.

Figure \ref{fig:realcompparams} summarizes the results of this fitting with a subplot for each free parameter. The $x$ axis of each subplot shows the range of values we modeled. The $y$ axis shows the projection of the combined $\chi^2$ matrix onto this parameter---that is, the minimum $\chi^2$ value found in the matrix while holding this parameter to the given value. This was then normalized to the $\chi^2$ value of the overall best fit.

\begin{figure}
\plotone{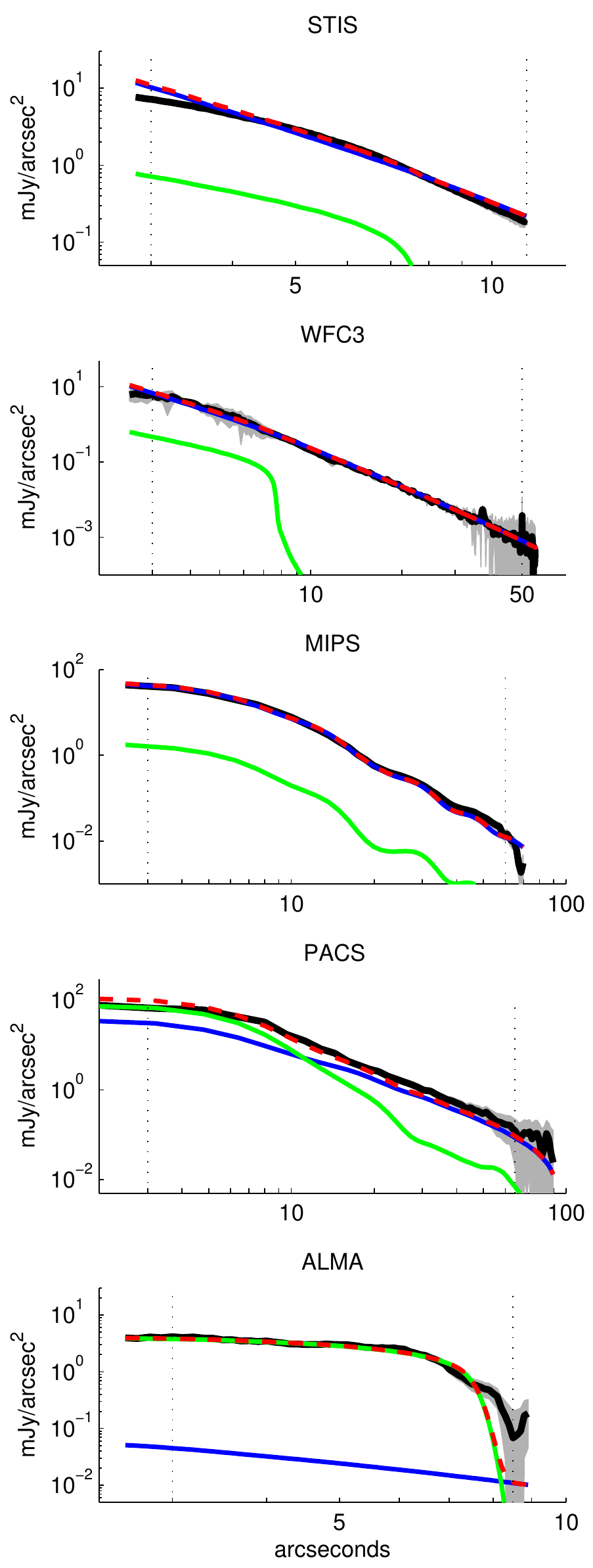}
\caption{\small The best fit model with a dust composition of 60\% astronomical silicates and 40\% refractory organics provides a good fit to all five data sets (NE side of the disk). The black lines are the data, the green lines are the parent body model, the blue lines are the halo model, and the dashed red lines are the total model. The vertical dashed lines show the range of data to which the model was fit.}
\label{fig:realcompfitsNE}
\end{figure}

\begin{figure}
\plotone{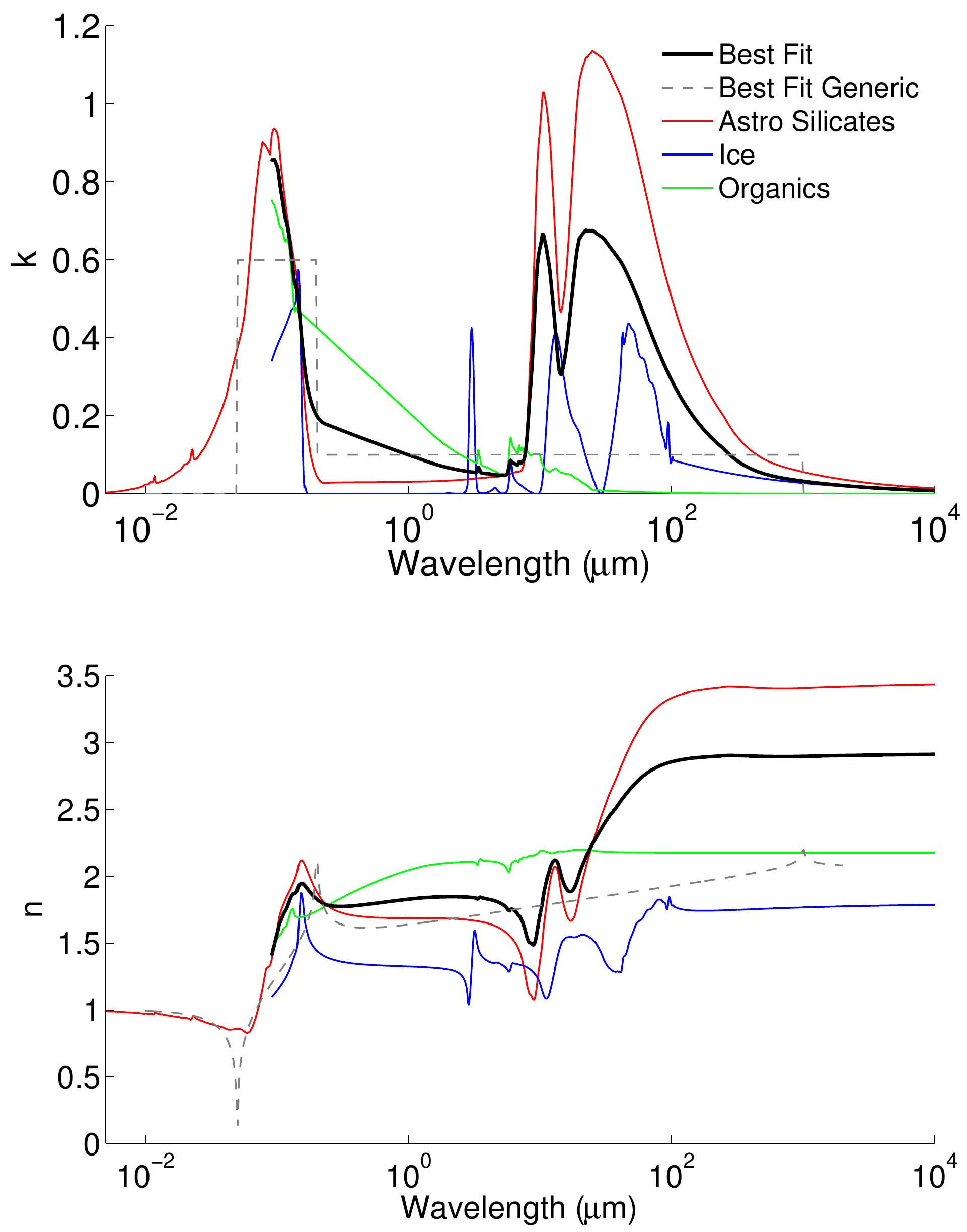}
\caption{\small The optical constants for our best fit model (60\% astronomical silicates and 40\% refractory organics), in addition to the optical constants of the three constituent materials we used and the best fitting generic constants.}
\label{fig:opticalconstantsNE}
\end{figure}

We found that a mixture of silicates and organics was preferred, while water ice and vacuum were not favored. The parameters of the best fit model are summarized in Table \ref{table:bestfit}. Figure \ref{fig:realcompfitsNE} shows that the best fit model radial profiles match all five data sets well. The optical constants for the best fitting model are shown in Figure \ref{fig:opticalconstantsNE}, along with the constants of the three constituent materials and the best fitting generic optical constants model we derived in \S \ref{sec:genericconstants}. We also provide the optical constants for our best fit composition in Table \ref{table:opticalconstants}.

\begin{deluxetable}{lc}
\tabletypesize{\scriptsize}
\tablewidth{0pt}
\tablecolumns{2}
\tablecaption{\small Properties of the Best Fit Model (of the NE side) \label{table:bestfit}}
\tablehead{\colhead{Parameter} & \colhead{Value for Best Model}}
\startdata
$r_\text{in}$ & 45 au \\
halo $r_\text{out}$ & 1800 au \\
PB $r_\text{out}$ & 150 au \\
halo $p$ & 2.4 \\
PB $p$ & 0.5 \\
halo $a_\text{min}$ & 0.1 $\micron$ \\
$a_\text{tran}$ & 5 $\micron$ \\
PB $a_\text{max}$ & 5000 $\micron$\tablenotemark{a} \\
halo $q$ & 3 \\
PB $q$ & 3.65 \\
$f_\text{sil}$ & 0.6 \\
$f_\text{ice}$ & 0 \\
$f_\text{org}$ & 0.4 \\
$f_\text{vac}$ & 0 \\
$M_\text{halo}$ & 1.13 $\times 10^{-2} M_\earth$ \\
$M_\text{PB}$ & 7.49 $\times 10^{-2} M_\earth$ \\
\enddata
\tablenotetext{a}{This value was fixed prior to fitting.}
\end{deluxetable}

\begin{deluxetable}{ccc}
\tabletypesize{\small}
\tablewidth{0pt}
\tablecolumns{3}
\tablecaption{\small Optical Constants of the Best Fit Model (60\% Astronomical Silicates, 40\% Organic Refractory Material) \label{table:opticalconstants}}
\tablehead{\colhead{$\lambda$ ($\micron$)} & \colhead{n} & \colhead{k}}
\startdata
0.091 & 1.4061 & 0.8523 \\ 
0.092 & 1.4268 & 0.8566 \\ 
0.093 & 1.4468 & 0.8577 \\ 
0.094 & 1.4667 & 0.8579 \\ 
0.096 & 1.4871 & 0.8574 \\ 
0.097 & 1.5092 & 0.8546 \\ 
0.098 & 1.5289 & 0.8502 \\ 
0.099 & 1.5474 & 0.8458 \\ 
0.100 & 1.5676 & 0.8408 \\ 
0.101 & 1.5888 & 0.8348 \\ 
\enddata
\tablecomments{Table \ref{table:opticalconstants} is published in its entirety in the electronic edition of the Astrophysical Journal. A portion is shown here for guidance regarding its form and content.}
\end{deluxetable}

Although we only used two or three values for each of the grain size parameters, our best fit model agreed well with the data, so trying additional values of the grain size parameters was not justified considering our aim was to constrain the composition. Furthermore, as shown by the thin colored curves in the top panels in Figure \ref{fig:realcompparams}, the general result for the grain composition---a mixture of astronomical silicates and organic refractory material with little to no ice---did not depend on the specific choice of grain size parameters. 

\subsection{SED}
\label{sec:sed}

For an important check on our best fit model, we compared it to the full-disk thermal SED of the disk at $\lambda \gtrsim 20$ $\micron$ where the flux from the outer components was dominant over the flux from the inner components.

To generate SED models, we first computed the dust temperature as a function of grain size and location from $a_\text{min}$ to $a_\text{max}$ and from $r_\text{in}$ to $r_\text{out}$ by equating the energy absorbed from stellar radiation (given by the model in \S\ref{sec:star}) with the emitted thermal energy. We computed the absorption efficiency, $Q_\text{abs}(\lambda,a)$, with the code \texttt{miex} \citep{wolf2004} using the given optical constants. The final SED was found by summing the contribution from grains of each size at each location, according to the model disk geometry and $n(r,a)$. The results of our procedure to generate SEDs agreed very well with the total flux in the thermal emission model images generated by \texttt{dustmap}. 

\begin{figure}
\plotone{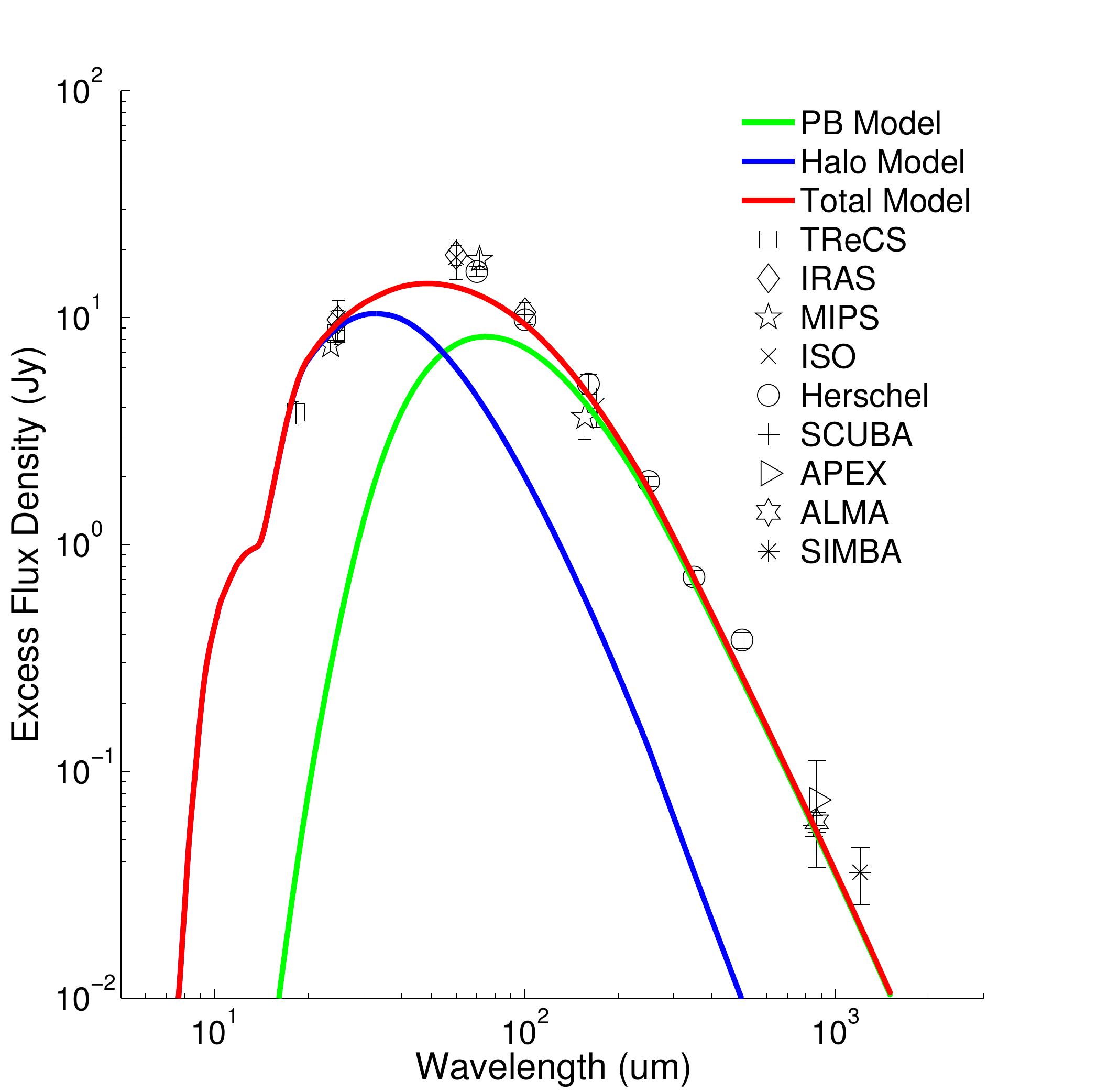}
\caption{\small The thermal SED of our best fit model, compared with the data given in Table \ref{table:SED}. The fit is good and provides a valuable confirmation of the model.}
\label{fig:realcompSEDNE}
\end{figure}

The best fit model SED is shown in Figure \ref{fig:realcompSEDNE}. Although our model slightly under-predicted the data at $\lambda \sim 70$ $\micron$, overall our model fit the data very well, supporting the application of the model at additional wavelengths.

The sub-mm slope of the SED is sensitive to the grain sizes of the parent body component. \citet{vandenbussche2010} examined the sub-mm slope of the $\beta$ Pic SED and concluded that the grain size distribution was shallower than predicted by a steady state collisional cascade. However, our best fit model has $q$ = 3.65 for the parent body component, as predicted for a collisional cascade.

In addition to the SED, we present three more comparisons with other data sets in the Appendix. These include the SW side of the disk (our fitting was only to the NE side), T-ReCS disk profiles in the mid-IR, and measurements of the disk's scattered light color. In all three cases, our model agrees satisfactorily with the additional data set.

\section{Discussion}
\label{sec:discussion}

\subsection{Sub-blowout Grains}

To find the blowout size predicted for the best fit model we found in \S \ref{sec:knownconstants}, we calculated the ratio of the radiation force to the gravitational force on a grain,
\begin{equation}
\label{eq:beta}
\beta = \frac{3L_\star}{16\pi GM_\star ac\rho}\frac{\int_0^\infty Q_\text{pr}(\lambda,a) F_{\lambda \star}(\lambda) \,\mathrm{d}\lambda}{\int_0^\infty F_{\lambda \star}(\lambda) \,\mathrm{d}\lambda},
\end{equation}
where $\rho$ is the grain density and $Q_\text{pr}(\lambda,a)$ is the radiation pressure efficiency for a grain of radius $a$ computed from the optical constants using the code \texttt{miex} \citep{wolf2004}. The blowout size occurs where $\beta$=0.5, with smaller grains (having larger $\beta$) being blown out. For the composition of our best fit model, the blowout size was 2.7 $\micron$, which was between our best fit $a_\text{min}$ and $a_\text{tran}$ values. That is, our best fit halo model consisted of a mixture of sub-blowout grains in the process of leaving the system plus barely bound grains on elliptical orbits.

One might expect that grains smaller than the blowout size would be depleted because they leave the system on short timescales. To test whether such a depletion was favored, we re-ran our fitting procedure (using the same mixture of common materials as in \S\ref{sec:knownconstants}) but with three dust components: a halo of sub-blowout grains, a halo of barely bound grains, and a parent body component. The spatial distributions of the two halo components were identical to each other and to the halo component used previously; the spatial distribution of the parent body component was also unchanged. The division between the sub-blowout and barely bound components was at the grain size where $\beta$=0.5 and the division between the barely bound halo and the parent body component was at the grain size where $\beta$=0.2. That is, $a_\text{max,sub}$ = $a_\text{min,barely}$ = $a$($\beta$=0.5), and $a_\text{max,barely}$ = $a_\text{min,PB}$ = $a$($\beta$=0.2). We used $a_\text{min,sub}$ = 0.1 $\micron$, $a_\text{max,PB}$ = 5000 $\micron$, $q$ = 3.65 for the parent body component, and $q$ = 3.0 for both halo components. The masses of the three components were free parameters in the fitting. The composition parameters were varied as before (from 0 to 1 in steps of 0.2). The grain sizes where $\beta$=0.5 and $\beta$=0.2 varied with the composition because different compositions have different $Q_\text{pr}(\lambda,a)$ values. The results were nearly the same as what we found in \S\ref{sec:knownconstants} with the best fit composition again $f_\text{sil}$ = 0.6, $f_\text{org}$ = 0.4. The grain size corresponding to $\beta$=0.2 was 6.4 $\micron$. Flux from the sub-blowout component was dominant in the scattered light bands and in thermal emission at 24 $\micron$; all three components contributed significantly at 70 $\micron$. The mass in sub-blowout grains was approximately the same as that in the original fitting where the halo component spanned both barely bound and sub-blowout grains in a single grain size distribution.

Next we tried forcing the model to be depleted in sub-blowout grains. Simulations by \citet{vanlieshout2014} predicted that the dust surface area per decade of grain size would be reduced by three orders of magnitude in sub-blowout grains compared to barely bound grains.\footnote{These simulations included grain sublimation, so the magnitude of the depletion and the precise $\beta$ value above which the depletion occurred depended on the orbital location of the dust. At 30 au (the location of the outer parent body belt in their model), the dust is unaffected by sublimation, and a depletion of three orders of magnitude in surface area per decade of grain size occurred at $\beta \simeq 0.5$ (see their Figure 5). This orbital location is most applicable to our situation, so we adopted these results.} We reran our three component fitting with this relative scaling between the sub-blowout and barely bound components imposed. We could not achieve a good fit to the data with any grain composition, confirming that sub-blowout grains were a necessary part of our model.    

Additional evidence exists for sub-blowout grains in the $\beta$ Pic system: detailed fits to the mid-IR spectral features used grains as small as 0.1 $\micron$ in size \citep{li1998,okamoto2004,li2012}. The data of \citet{okamoto2004} are particularly significant since they see spatially distributed spectral features from sub-blowout crystalline and amorphous silicates to 30 au, where it appears the features become lost in the noise. \citet{devries2012} fit olivine features at 34 and 69 $\micron$ with a model emphasizing grain sizes of 1--3 $\micron$ and with the grain placement consistent with the inner part of the parent body disk, again showing the importance of sub-blowout grains in the overall SED.  Finally, the SED of $\beta$ Pic (Figure \ref{fig:realcompSEDNE}) shows its 24 $\micron$ flux density to be within a factor of two of the peak, whereas it is more typical of debris disks to have a difference of an order of magnitude. That is, the warm spectrum arising from small grains is unusually prominent compared with typical disks. Taken together with our models that showed no significant discontinuity could be tolerated in the grain size distribution at the blowout size, these observations support our assumption that small grains in the halo---many of them below the blowout size---are dominant in the output at wavelengths shorter than 50 $\micron$. The emission at longer wavelengths is then contributed primarily by the larger grains in the parent body ring (created in the collisional cascade therein), as required to fit the well-resolved image with ALMA.

Sub-blowout grains have been inferred from the modeling of other young, bright debris disks. For example, when modeling the debris disk around Fomalhaut, \citet{acke2012} found that a significant contribution to the observed flux came from sub-blowout grains. \citet{rodigas2015} used $a_\text{min}$ = 0.1 to fit the thermal and scattered light of the HR 4796A disk.

\subsection{The Dust Composition}

Here we compare the composition/optical constants of our best fit model with those found in other studies. We reiterate that many models of debris disks simply assumed the dust was composed of astronomical silicates \citep[e.g.][]{Krist2010,golimowski2011,ertel2011}. Other studies that did constrain the optical constants relied on fitting the thermal SED only and did not match the brightness of the disk in scattered light \citep[e.g.][]{lebreton2012,donaldson2013}.

\citet{li1998} modeled the SED of $\beta$ Pic with a focus on fitting the detailed shape of the $\sim$10 $\micron$ feature, so their constraints were strongest for the inner dust component. They found that two grain populations were needed: silicate grains with organic refractory mantles and crystalline silicate grains. The silicate grains with organic refractory mantles were roughly consistent with the composition we found for the dust in the outer disk. \citet{okamoto2004} and \citet{li2012} found that the crystalline component detected near 10 $\micron$ was concentrated in the inner disk, while \citet{devries2012} detected crystalline grains via their 34 and 69 $\micron$ features in the outer disk. In both cases, the crystalline materials account for only a few percent of the dust mass, but are readily detectable in these small amounts because of their sharp spectral features. Both because of their small concentration and because the broad spectral character of the crystalline material is similar to that of the amorphous material assumed in our model \citep[e.g.][]{fabian2000}, basing the model on the amorphous material is acceptable given our emphasis on providing as simple a fit as possible.

\citet{min2011} derived expected dust compositions based on the solar elemental abundances, yielding four species with the following range of mass fractions: silicates (24--47\%), FeS (7--14\%), carbonaceous dust (0-20\%), and water ice (39--49\%). The range is due to the unknown fraction of carbon that ended up in dust versus CO gas. This mix of compositions was used to successfully fit the \textit{Herschel} thermal images of the Fomalhaut disk \citep{acke2012}, but was not quantitatively compared to the scattered light observations.

Dust particles in the Uranus ring system are very dark in scattered light and lack water ice features \citep{karkoschka2001}, making them potential analogs for the dust in the $\beta$ Pic disk. Other solar system particles, like those in the rings of Saturn, do not share these properties, however.

The most direct comparison with our work is the characterization of the composition of the HR 4796A debris disk using both scattered light and thermal emission by \citet{rodigas2015}. One of their best fitting models was an isolated case involving a large fraction of metallic iron. We did not include iron in our fitting of $\beta$ Pic, but we consider it unlikely that the dust grains contain much metallic iron unless they have been exposed to very high temperatures. This fit illustrates our argument that multiple types of material are in principle capable of producing the optical constants needed to fit debris disk behavior. Excluding this case, \citet{rodigas2015} found that silicates and organics were generally preferred and water ice was not. This agrees with our findings for $\beta$ Pic and suggests that there may be some commonality to the composition of different debris disks.    

\section{Summary}
\label{sec:conclusion}

Matching the thermal emission and scattered light data simultaneously has been a persistent problem for debris disk modeling. Here we investigated whether this problem could be solved by varying the optical constants (and thus composition) of the debris disk dust. We tested this on the $\beta$ Pic disk, for which there are high-quality well-resolved images at many wavelengths, including in both scattered light and thermal emission. We fit our models to data from five instruments: \textit{HST}/STIS, \textit{HST}/WFC3, \textit{Spitzer}/MIPS, \textit{Herschel}/PACS, and ALMA. The main results of our modeling were as follows:
\begin{itemize}
  \item When assuming the dust was composed entirely of astronomical silicates, we could not achieve a successful fit. This resulted in a model that was too bright in scattered light relative to its thermal emission---the same offset found by studies that attempted to model other debris disks using only astronomical silicates.
  \item We found that a generic model for the optical constants with only a few free parameters could achieve a much-improved fit. This demonstrated that varying the optical constants was capable of solving the problem.
  \item Since a variety of materials might be capable of yielding the necessary optical constants, other constraints must be used to narrow the selection of grain compositions.  
  \item We modeled the dust as a combination of plausible materials---astronomical silicates, water ice, refractory organics, and vacuum. We found that a good fit could be achieved with a mix of silicates and organics, and that ice and vacuum were not favored.
  \item This model also reproduced well the observed thermal SED, the scattered light colors, and the images from T-ReCS at two mid-IR bands.
  \item The resulting best fit composition was similar to candidates for the composition of the HR 4796A debris disk found also by simultaneously fitting the thermal and scattered light observations \citep{rodigas2015}.
\end{itemize}
With continued observations from \textit{HST} and ALMA and future observations from \textit{JWST}, the number of debris disks with high-quality data across the electromagnetic spectrum will grow. The composition of these disks can be measured by the method described here. This will allow compositions to be determined and compared for many debris disks.

\acknowledgments
We thank Daniel Apai and Glenn Schneider for sharing their STIS data and providing useful scientific advice. We thank William Dent for sharing the ALMA data, and Charles Telesco for sharing the T-ReCS data. We thank Chris Stark for providing tech support for \texttt{dustmap}. We thank Aigen Li for sharing his tabulated optical constants for water ice and refractory organics. We also thank the referee for providing useful feedback. This research made use of Tiny Tim/\textit{Spitzer}, developed by John Krist for the Spitzer Science Center. The Center is managed by the California Institute of Technology under a contract with NASA. This work was supported by NASA grants NNX13AD82G and 1255094.

{\it Facilities:} \textit{HST} (WFC3, STIS), \textit{Spitzer} (MIPS), \textit{Herschel} (PACS), ALMA, Gemini:South (T-ReCS).

{\it Software:} MATLAB, Dustmap, IDL, MIPS Data Analysis Tool, TinyTim, STinyTim.

\appendix
\section{Comparison with Additional Data}
\label{sec:appendix}

\subsection{The SW Side}
\label{sec:swcheck}

Our constraints on the dust composition used data only from the NE side of the disk. Here we use the data from the SW side of the disk as a check on our results. We generated model images using the same grain size and composition parameters as our best fit model to the NE side, but with the spatial parameters found in \S \ref{sec:spatial} for the SW side (parent body $r_\text{out}$ = 155 au instead of 150 au, and more significantly the halo $p$ = 3.1 instead of 2.4).

\begin{figure}
\plotone{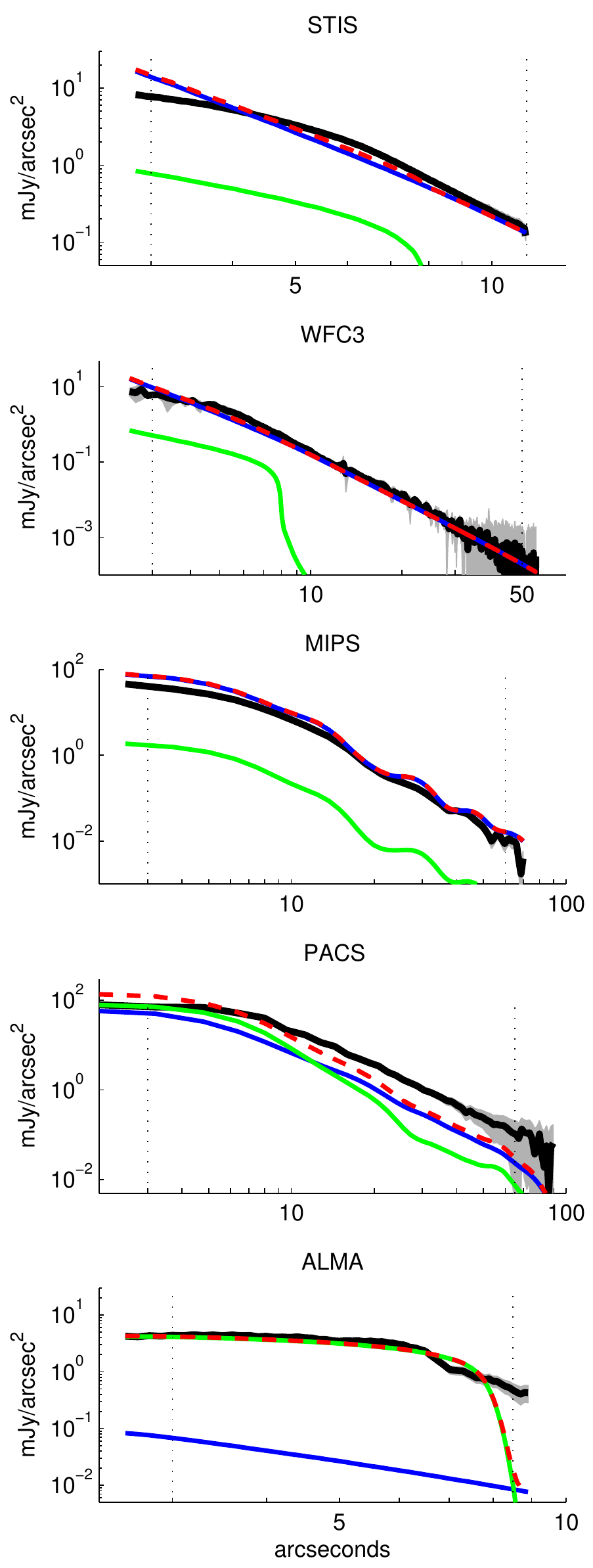}
\caption{\small Models generated with the grain properties derived from fits to the NE side but with the spatial parameters found for the SW side. The models were fit to the data from the SW side of the disk. The halo component matched the \textit{HST} and MIPS bands well, but was too faint compared to the 70 $\micron$ PACS data. In addition, the parent body component fit the ALMA data well, but contributed too much at 24 and 70 $\micron$. The black lines are the data, the green lines are the parent body model, the blue lines are the halo model, and the dashed red lines are the total model. The vertical dashed lines show the range of data to which the model was fit.}
\label{fig:realcompfitsSW}
\end{figure}

The results are shown in Figure \ref{fig:realcompfitsSW}. The models fit the STIS, WFC3, and MIPS data well, but the halo component under-predicts the observed flux at 70 $\micron$. The parent body model fits the ALMA data well, but is somewhat too bright in the MIPS and PACS bands. The masses of the model components found by this fitting were $M_\text{halo} = 6.01 \times 10^{-3} M_\earth$ and $M_\text{PB} = 8.60 \times 10^{-2} M_\earth$. Compared to the NE side, the SW side had a less massive halo and a more massive parent body component.

Our fitting to the the SW side implicitly assumed that the known asymmetry between the two sides was caused by differences in the spatial distribution of the dust, rather than differences in the grain properties. Perhaps the fit to the SW side could be improved by using different grain size parameters. The SW side also hosts a large clump seen in thermal emission at several wavelengths and in CO gas \citep{telesco2005,dent2014}, which may be the site of a recent massive collision and may contribute to the asymmetry. A detailed study of the differences between the NE and SW sides of the disk, however, is beyond the scope of this paper.  

\subsection{Gemini/T-ReCS}
\label{sec:trecs}

$\beta$ Pic was imaged with Gemini/T-ReCS in five bands: 8.7, 11.7, 12.3, 18.3, and 24.6 $\micron$, and these data were published by \citet{telesco2005}. Here we used the images at 18.3 and 24.6 $\micron$, wavelengths at which the outer disk components contributed significantly. We obtained the rotated, calibrated images with units of mJy/pixel. From each image we subtracted a constant background value, derived using the IDL program \texttt{mmm.pro}. These values were 0.005 $\pm$ 0.225 mJy/pixel at 18.3 $\micron$ and 0.214 $\pm$ 1.299 mJy/pixel at 24.6 $\micron$. We converted the surface brightness to units of mJy arcsec$^{-2}$ using a pixel size of 0\farcs09 for the 18.3 $\micron$ image and 0\farcs086 for the 24.6 $\micron$ image. We smoothed each image with a boxcar kernel roughly the size of the FWHM of the instrument PSF (5$\times$5 and 7$\times$7 pixels for the 18.3 and 24.6 $\micron$ images, respectively).

\begin{figure}
\plotone{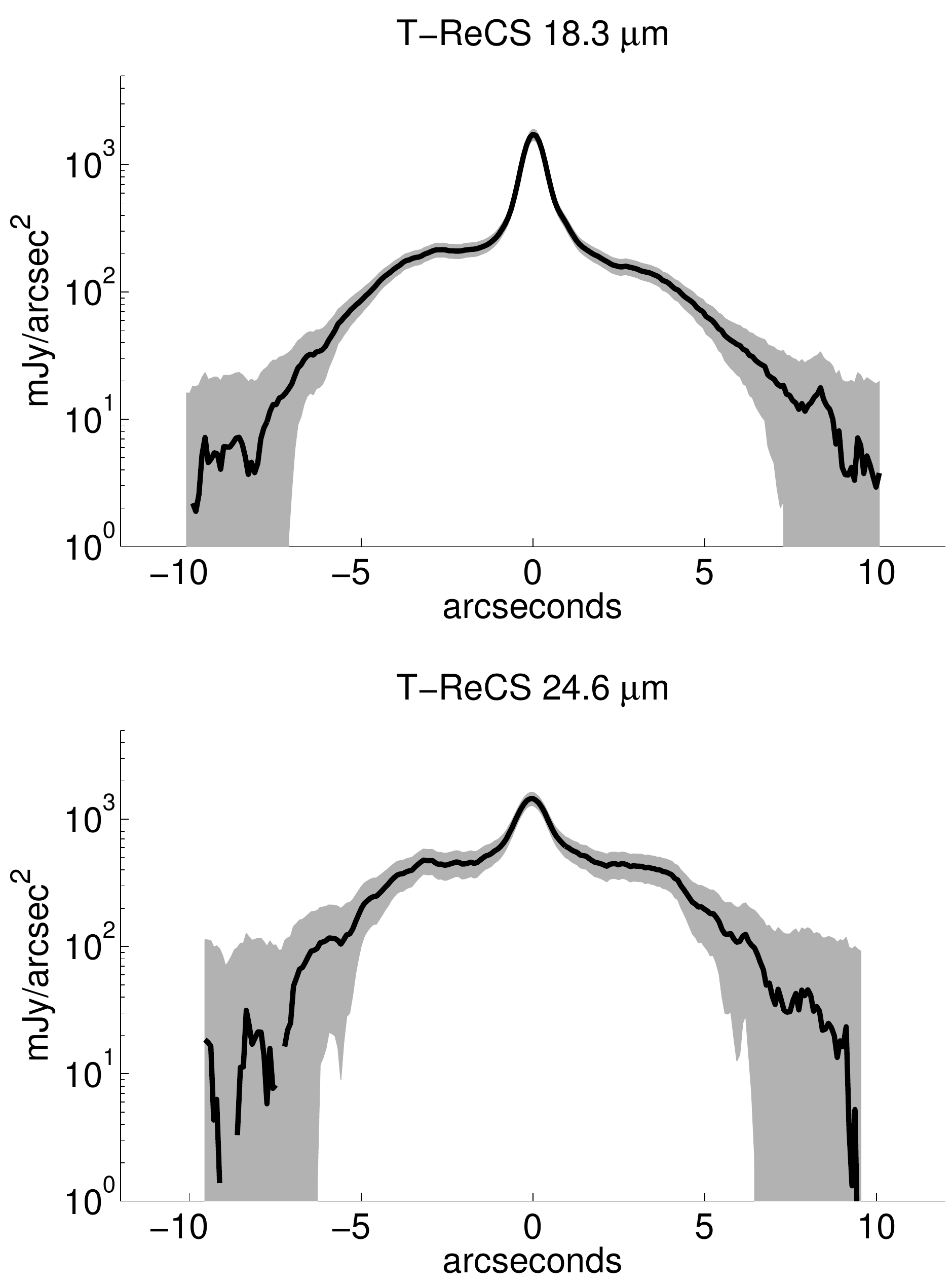}
\caption{\small Profiles of the T-REcS images of the $\beta$ Pic disk at 18.3 and 24.6 $\micron$. The gray region is the uncertainty along the profiles. The NE side of the disk is to the right, the SW side is to the left. The central peak is the unresolved flux from the central star and the inner disk component. Outside of that, the broad shoulder is the flux from the outer disk components.}
\label{fig:TRECSProfiles}
\end{figure}

We extracted radial profiles using a cut width of 9 pixels (0\farcs81 at 18.3 $\micron$, 0\farcs774 at 24.6 $\micron$). The profiles are shown in Figure \ref{fig:TRECSProfiles}. The uncertainty on the profiles was the combination in quadrature of 10\% calibration uncertainty and the mJy/pixel uncertainty of 0.13 and 0.7 for the 18.3 and 24.6 $\micron$ images, respectively, from Table 1 of \citet{telesco2005}. The profiles had a central, unresolved component arising from the star plus the warm inner disk component with the flux outside of this arising from the outer disk components.

We included the photometry measurements of the whole disk as given by \citet{telesco2005} for these data in our Table \ref{table:SED}. The T-ReCS flux at 24.6 $\micron$ was higher than the MIPS flux and ISO flux at similar wavelengths, which may be due to a calibration problem (this is supported by the relatively large background value we found for the 24.6 $\micron$ image).

\begin{figure}
\plotone{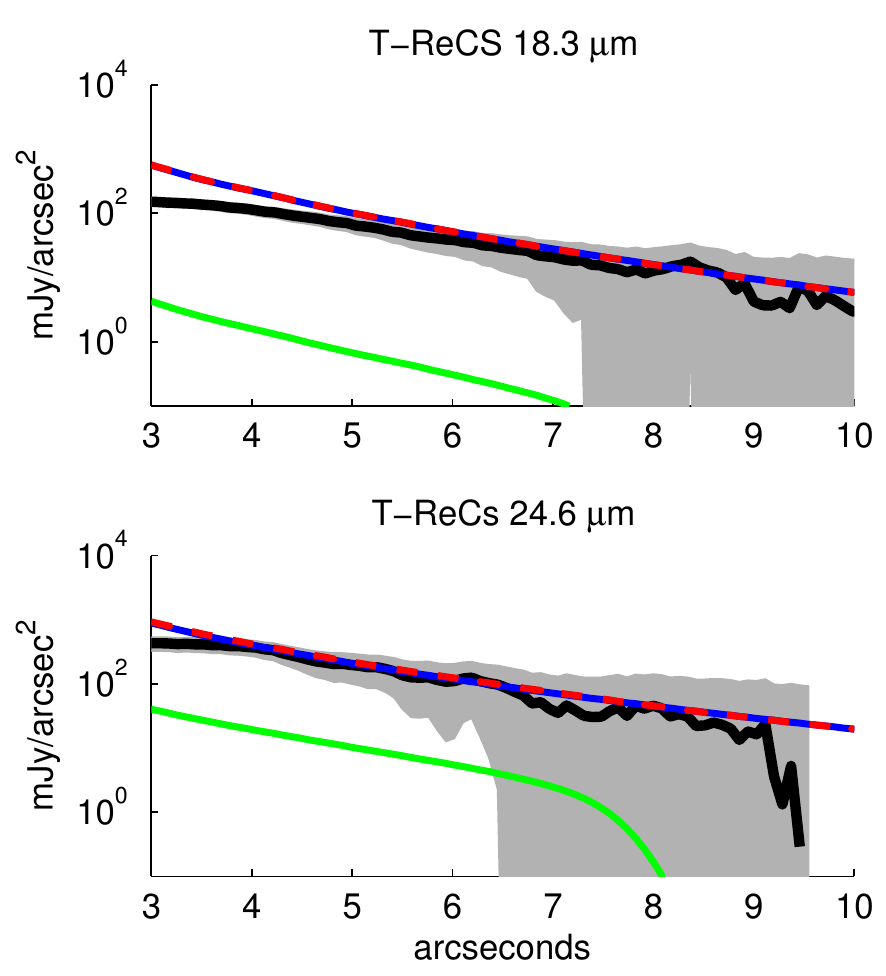}
\caption{\small The best fit model compared with NE side profiles of the T-ReCS data at 18.3 and 24.6 $\micron$. We achieved a good fit with the exception of the shape of the model at 18.3 $\micron$ at small $r$.}
\label{fig:realcomptrecsNE}
\end{figure}

We compared our best fit model with the outer parts of the 18.3 and 24.6 $\micron$ T-ReCS profiles. We convolved model images at these wavelengths with PSFs that were modeled as symmetric 2D Gaussians with FWHM of 0\farcs54, and 0\farcs72 for the 18.3 and 24.6 $\micron$ images, respectively. The comparison is shown in Figure \ref{fig:realcomptrecsNE}. We find good agreement between our models and these data. The 18.3 $\micron$ image is the shortest wavelength in the thermal regime at which our model was tested.

\subsection{Scattered Light Color}

\citet{golimowski2006} imaged $\beta$ Pic's disk with the \textit{HST}/ACS High Resolution Channel in three scattered light bands: F435W, F606W, and F814W with central wavelengths 0.4311, 0.5888, and 0.8115 $\micron$, respectively. They found that the disk was redder than the star, and that the disk became somewhat redder moving outwards along the midplane. Specifically, for their PSF-deconvolved images, the F435W$-$F606W color ranged from 0.1 to 0.2, and the F435W$-$F814W color ranged from 0.2 to 0.35 along the disk (see their Figure 18).

\citet{golimowski2006} investigated whether they could constrain the dust composition and minimum grain size using their measured scattered light colors. They found that many combinations of parameters could fit their data, which supports the premise of our work---that both scattered light and thermal data are required to constrain the composition. They did, however, exclude very porous grains (90\%), which always resulted in scattered light colors bluer than the star. 

\begin{figure}
\plotone{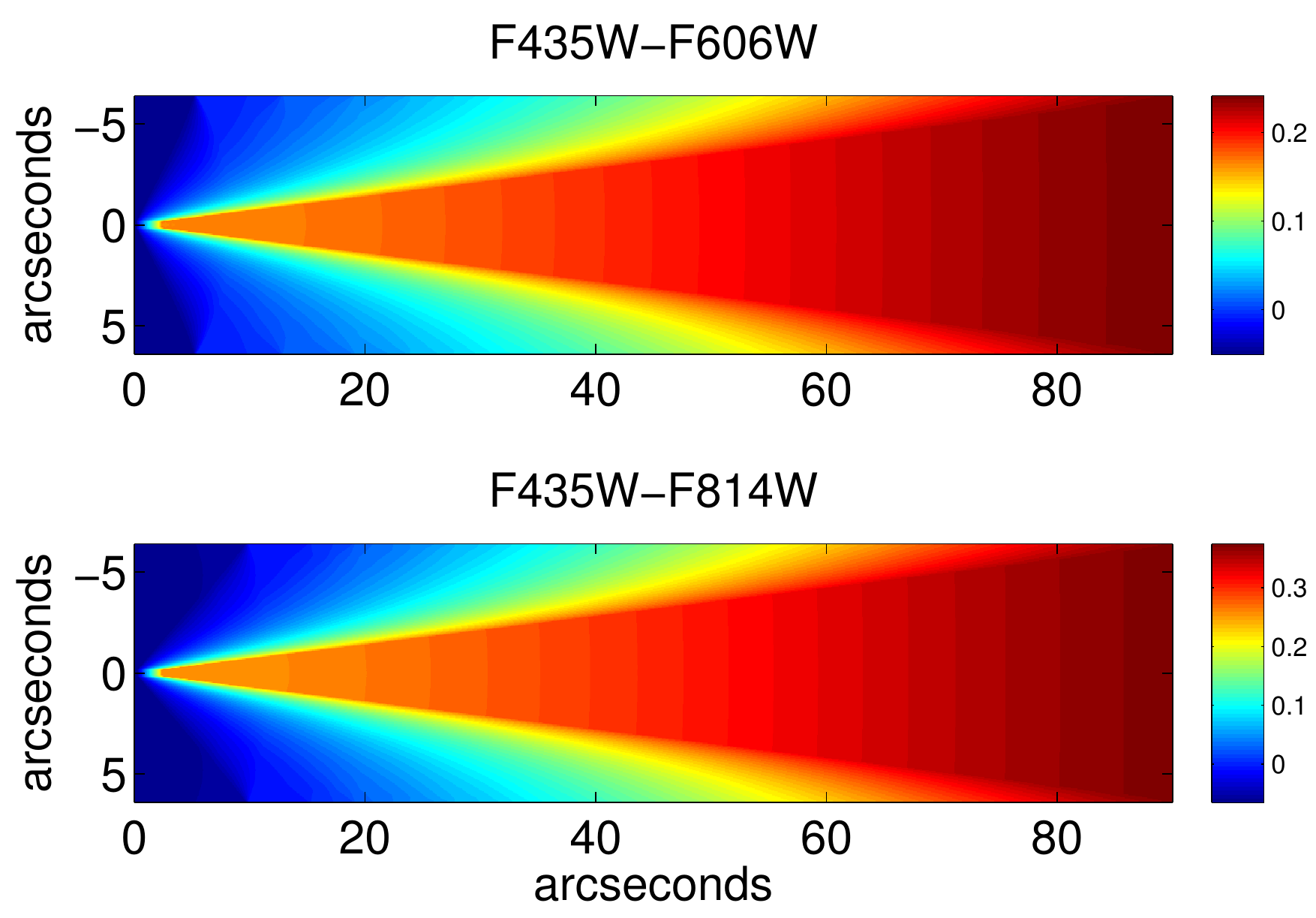}
\caption{\small Images of the F435W-F606W and F435W-F814W colors of our model disk. The disk is bluer in regions dominated by forward- and back-scattering.}
\label{fig:ScatteredColorImages}
\end{figure}

We generated model images at these ACS wavelengths to see if our best fit halo component (which dominated the scattered light signal) showed a similar behavior in its scattered light colors. Figure \ref{fig:ScatteredColorImages} shows the F435W$-$F606W and F435W$-$F814W colors of our model. To make these images, we divided the model image at each wavelength by the flux density of the star at that wavelength (as discussed in \S \ref{sec:star}) and then divided the two images by each other and converted the result to a magnitude scale. Because we were comparing with the deconvolved ACS data, we did not convolve our model images with any model PSF. Our results generally agreed with the ACS data---the midplane of the disk was redder than the star by a couple of tenths of magnitude, and became redder farther from the star.

Our model used the same grain sizes and composition at all locations in the disk, so the change in disk color across the image must result from the wavelength dependence of the scattering phase function. Interior to $r_\text{in}$ and also above the disk wedge---locations dominated by forward- or back-scattering---the dust was bluer (at some points even bluer than the star), whereas in the disk plane where scattering occurred at angles closer to $\sim$90$^\circ$, the dust was red. The increasing redness of the disk outwards along the midplane likely also arises due to an increasing proportion of the scattering happening at $\sim$90$^\circ$.

\begin{figure}
\epsscale{1.0}
\plotone{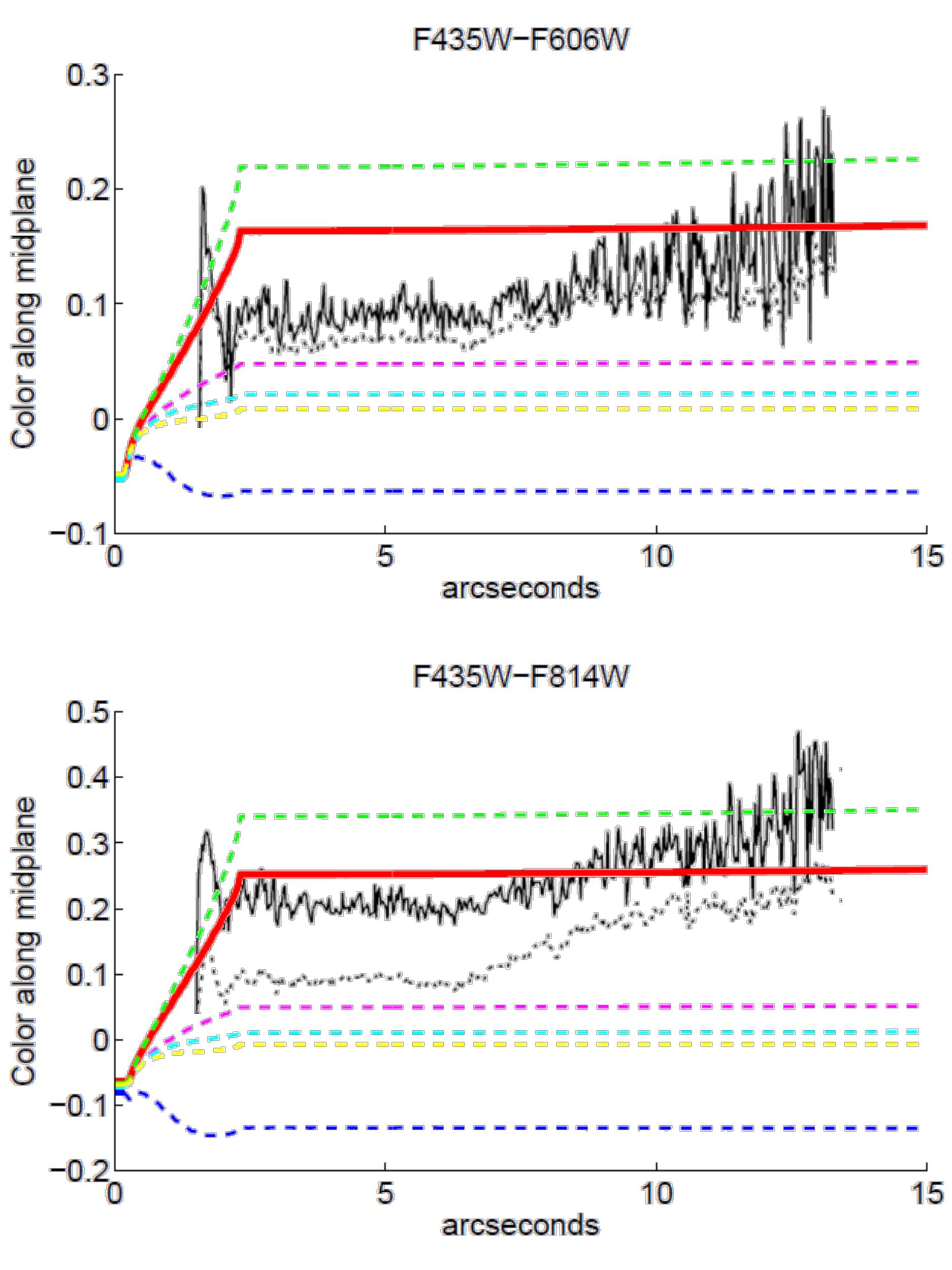}
\caption{\small The F435W-F606W and F435W-F814W colors of the disk (relative to the star) along the disk's miplane. The black line is the measured color profile from \citet{golimowski2006} (the dotted line is their measurement prior to deconvolution). The colored lines are our model with various dust compositions: red is $f_\text{sil}$ = 0.6 and $f_\text{org}$ = 0.4 (our best fit model), green is $f_\text{org}$ = 1, magenta is $f_\text{sil}$ = 1, blue is $f_\text{ice}$ = 1, cyan is $f_\text{sil}$ = 0.6 and $f_\text{ice}$ = 0.4, and yellow is $f_\text{sil}$ = 0.6 and $f_\text{vac}$ = 0.4. There is relatively good agreement between the data and our best fit model, especially for F435W-F814W. The data for both colors falls between the silicates and organics models (implying a mixture of the two materials) and is redder than mixtures with significant amounts of water ice or vacuum, which agrees with the results of our fitting. The bluer region of the disk inside $r<2.3\arcsec$ is from within $r_\text{in}$ of our model, so the flux is starlight that was highly forward- or back-scattered. Our model's color profile was fairly constant out to 15$\arcsec$ shown here, but does get redder toward the outer edge.}
\label{fig:ScatteredColorCuts}
\end{figure}

In Figure \ref{fig:ScatteredColorCuts} we plot profiles of the color images along the midplane as well as the color profile measured by \citet{golimowski2006}. They smoothed their image with a 7$\times$7 pixel boxcar before extracting a profile along the midplane, so we extracted a four pixel wide profile to capture approximately the same region of the image (the ACS pixels were half the size of our model pixels). This confirmed the general agreement of our model with these measurements. In the range of the radial profile where \citet{golimowski2006} measured the disk colors (3\arcsec--13\arcsec), our model color profiles show a constant color. However, using a wider profile cut to generate the profiles would result in an increasing red color over this range of the profile, because less of the bluer flux from above the wedge would be included with increasing distance from the star.

\begin{figure}
\plotone{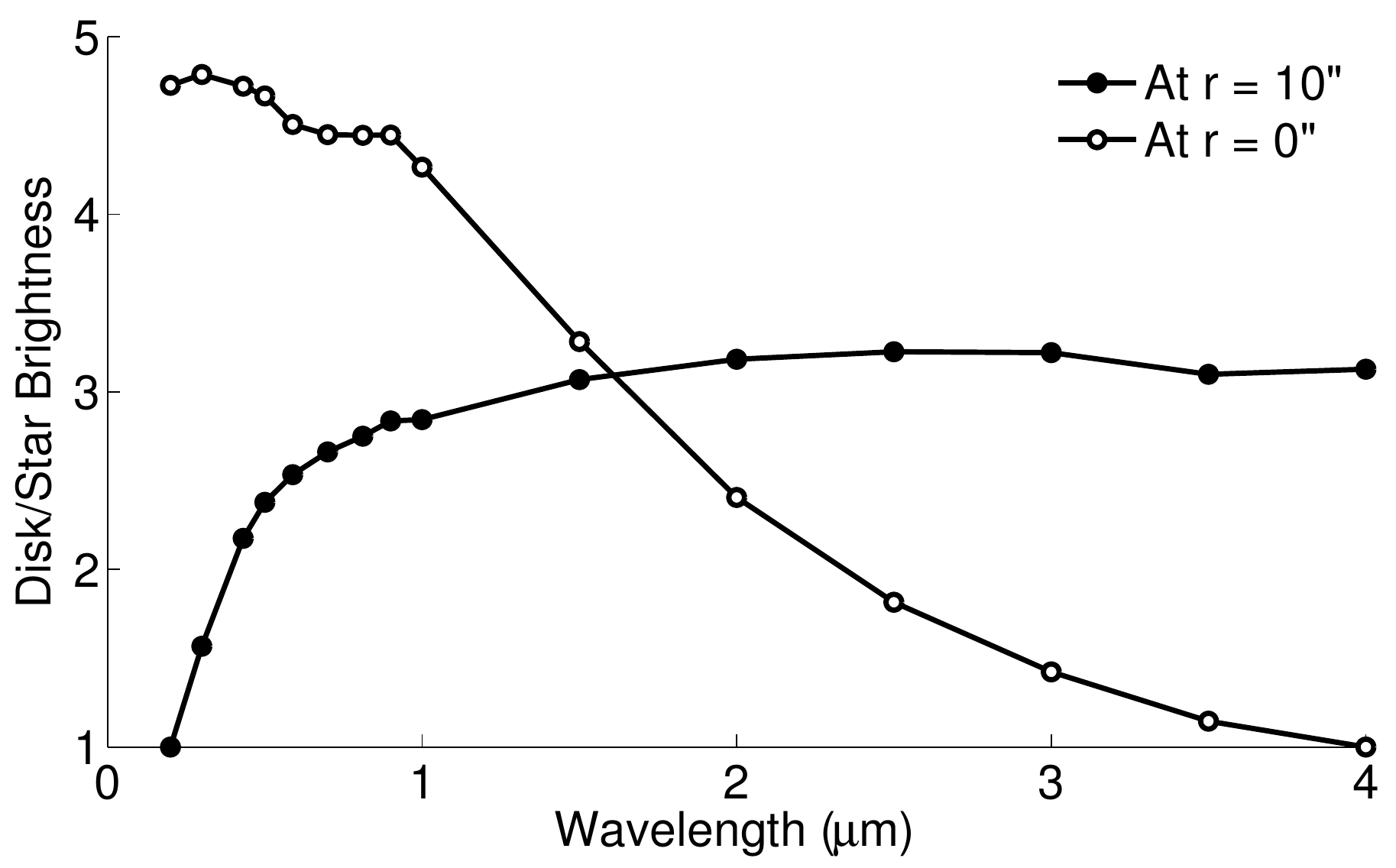}
\caption{\small The SED of the scattered light (relative to the SED of the star) at two locations in our model image: at the origin (where the light is highly forward- and back-scattered) and in the midplane of the disk 10$\arcsec$ from the star. The color at the origin was blue, where the color at $r=10\arcsec$ was red. The y axis units are arbitrary (normalized to the minimum value of each curve).}
\label{fig:ScatteredColorSED}
\end{figure}

Next we looked into the color of the dust predicted by our model at wavelengths beyond those measured by ACS, STIS, and WFC3. We generated models of the halo component from 0.2--4 $\micron$, and normalized them by the brightness of the star at those wavelengths. In Figure \ref{fig:ScatteredColorSED} we show the resulting scattered light SEDs extracted at the origin and at $r=10\arcsec$ on the disk midplane. The SED from the disk midplane showed the dust reddening across the visible, but the color became more neutral at longer wavelengths. At the origin, which probed only the forward- and back-scattered light, the dust was blue across this whole wavelength range, although the gradient of the color was shallower at longer wavelengths.

\bibliographystyle{aasjournal}

\end{document}